\documentclass[11pt,journal,draftcls,onecolumn,peerreviewca]{IEEEtran}
\usepackage{lineno}
\usepackage{multirow}
\usepackage{amsfonts}
\usepackage{amsmath}
\usepackage{amssymb}
\usepackage{diagbox}
\usepackage{bm}
\usepackage{xcolor,graphicx,float}
\usepackage{color, soul}
\usepackage{stfloats}
\usepackage[numbers,sort&compress]{natbib}
\usepackage[amsmath,thmmarks]{ntheorem}
\usepackage{theorem}
\usepackage{algorithm}
\usepackage{algorithmic}
\usepackage{graphicx}
\usepackage{subfigure}
\usepackage{pict2e}
\usepackage{ulem}

\theoremheaderfont{\sc}\theorembodyfont{\upshape}
\theoremstyle{nonumberplain}
\theoremseparator{}
\theoremsymbol{\rule{1ex}{1ex}}

\begin{document}
\title{Grid-less Variational Direction of Arrival Estimation in Heteroscedastic Noise Environment}

\author{Qi~Zhang, Jiang~Zhu, Yuantao Gu and Zhiwei~Xu\thanks{Qi Zhang, Jiang Zhu and Zhiwei Xu are with the Ocean College, Zhejiang University, Zhoushan, CHINA, email: \{zhangqi13, jiangzhu16, xuzw\}@zju.edu.cn. Yuantao Gu is with Department of Electronic Engineering, Tsinghua University, Beijing, China, email: gyt@tsinghua.edu.cn. }
}

\maketitle
\begin{abstract}
Horizontal line arrays are often employed in underwater environments to estimate the direction of arrival (DOA) of a weak signal. Conventional beamforming (CB) is robust but has wide beamwidths and high-level sidelobes. High-resolution methods such as minimum-variance distortionless response (MVDR) and subspace-based MUSIC algorithm, produce low sidelobe levels and narrow beamwidths, but are sensitive to signal mismatch and require many snapshots and the knowledge of number of sources. In addition, heteroscedastic noise where the variance varies across observations and sensors due to nonstationary environments degrades the conventional methods significantly. This paper studies DOA in heteroscedastic noise (HN) environment, where the variance of noise is varied across the snapshots and the antennas. By treating the DOAs as random variables and the nuisance parameters of the noise variance different across the snapshots and the antennas, multi-snapshot variational line spectral estimation (MVALSE) dealing with heteroscedastic noise (MVHN) is proposed, which automatically estimates the noise variance, nuisance parameters of the prior distribution, number of sources, and provides the uncertain degrees of DOA estimates. When the noise variance only varies across the snapshots or the antennas, the variants of MVHN, i.e., MVHN-S and MVHN-A, can be naturally developed. Finally, substantial numerical experiments are conducted to illustrate the proposed algorithms' performance, including a real data set in DOA application.
\end{abstract}
{\bf keywords}: Variational Bayesian inference, von Mises distribution, grid-less, DOA estimation, heteroscedastic noise
\section{Introduction}\label{Intro}
Horizontal line arrays (HLA) are often used for target bearing estimation based on the direction of arrival (DOA) of the received signal. For the far-field scenario, the distance between the source and the receiver is great compared with the array aperture. The signals are assumed to arrive as plane waves. A popular method of array signal processing is conventional beamforming (CBF), in which the received signals are delayed and summed based on the sensor configuration relative to the signal look direction. When the steering angle matches the DOA of a given target, the target signals are coherently summed. In contrast, noise between the receivers is uncorrelated and is summed incoherently. As a consequence, the signal to noise ratio (SNR) at the output of the beamformer is enhanced, and the SNR gain compared to the input SNR is termed as the array gain (AG).

The advantage of CBF, as is well known, is the robustness against signal mismatch between the assumed and the actual signal wavefront. The disadvantages of CBF are the wide beamwidths, which make it difficult to detect close targets within the same beam, and the high sidelobe levels, which makes it hard to detect the weak source in the presence of a loud interferer. To overcome the deficit of CBF, high-resolution methods such as minimum-variance distortionless response (MVDR) and subspace-based MUSIC algorithm are proposed \cite{Trees}. These algorithms yield low sidelobe levels and narrow beamwidths, while they are sensitive to the signal mismatches at the same time. Besides, all the methods require knowledge of the number of sources.

Often, the main assumption used in most DOA problem is that noise is wide sense stationary, i.e., the noise variance is constant across both antennas and snapshots. However, for long observation times, the noise may be heteroscedastic, i.e., the noise variance changes either in antennas or snapshots, or both. Thus, developing algorithms taking noise variation into account will be beneficial for DOA. Mathematically, the modelings of the noise variance can be classified into four cases \cite{PeterHN}:
\begin{description}
\item[Case $\rm \uppercase\expandafter{\romannumeral1}$:] \quad Noise is wide-sense stationary in both antennas and snapshots.
\item[Case $\rm \uppercase\expandafter{\romannumeral2}$:] \quad Noise is wide-sense stationary only in antennas, and is varied across the snapshots.
\item[Case $\rm \uppercase\expandafter{\romannumeral3}$:] \quad Noise is wide-sense stationary only in snapshots, and is varied across the antennas.
\item[Case $\rm \uppercase\expandafter{\romannumeral4}$:] \quad Noise is heteroscedastic across both antennas and snapshots.
\end{description}
Case $\rm \uppercase\expandafter{\romannumeral1}$ corresponding to wide-sense stationary noise in both antennas and snapshots is the most widely used assumption, and many well-known methods are developed. For the other three cases,  practical scenarios also exist. For example, for noise Case $\rm \uppercase\expandafter{\romannumeral2}$, underwater acoustic channel is time-varying, and it varies in seasons, areas and situation of sea surface. In addition, spatial movement of the source changes in the propagation conditions such as sound speed profile, which lead to channel variation. As a result, one can model the additive noise with time-varying variance \cite{AmiriICASSP, AmiriEURASIP}. For noise Case $\rm \uppercase\expandafter{\romannumeral3}$, it is also known as nonuniform noise which models the sensors with hardware nonidealities in receiving channels
\cite{Nickel}, as well as for arrays with position dependent noise (for example, hydrophones near to the surface have a larger noise variance because of passing ocean wave on the surface). As for noise Case $\rm \uppercase\expandafter{\romannumeral4}$, it may correspond to the situation where a moving target/interference with high bearing rate dominates a few nearby sensors (spatially varying noise) \cite{Bagg, Riahi}. It is worth noting that in practice, after the array was deployed and when the target is absent, the measured data sampling over both time and space can be used to decide which noise modeling is more appropriate.

Most works focus on studying the DOA under noise Case $\rm \uppercase\expandafter{\romannumeral1}$ assumption. Since the sources are sparse in the spatial domain, many compressed sensing-based DOA estimation methods are proposed and can be classified into three cases: on-grid methods, off-grid methods, and grid-less methods \cite{yangzaibook}. On-grid methods refer to discretize the DOAs belonging to $[-90,90]^{\circ}$ into a number of grids. Off-grid methods are based on the on-grid approaches with an additional grid refinement. Grid-less methods treat the frequency as the continuous parameter and completely overcome the model mismatch \cite{Chi}, compared to on-grid and off-grid methods, thus have attracted much more attention in recent years. Specifically, grid-less methods can be classified into two categories, one is the high complexity atomic norm-based approaches which involve solving a semidefinite programming \cite{Yang2, Yang3}, the other is the Bayesian approach which treats the direction of arrivals as random variables \cite{VALSE}. It is also worth noting that both CBF and MVDR are on-grid methods, while root MUSIC is grid-less.

It has already been shown that traditional DOA estimation methods under Case $\rm \uppercase\expandafter{\romannumeral1}$ noise assumption are sensitive to the noise models \cite{Sinath}, and their performance degrades significantly when the assumption is not met \cite{FuVaccaro}. For example, the number of effective samples for constructing the covariance matrix can be relatively small and leads to the so-called snapshot deficient condition \cite{Bagg}. As a result, several algorithms have been proposed to tackle the DOA problem under the three noise modeling cases \cite{PeterRobust, Gershman, Kung, GuoDOA, PeterHN}. For Case $\rm \uppercase\expandafter{\romannumeral2}$, a sparse Bayesian learning (SBL) algorithm robust to modest mismatch in simulations is proposed, and improved source localization performance is demonstrated with experimental data \cite{PeterRobust}. The deterministic maximum likelihood (ML) estimator \cite{Gershman} and stochastic ML estimator \cite{Kung} are proposed for the nonuniform white noise with an arbitrary diagonal covariance matrix as Case $\rm \uppercase\expandafter{\romannumeral3}$. In \cite{PeterHN}, Case $\rm \uppercase\expandafter{\romannumeral4}$ is studied, and on-grid SBL is proposed to estimate the heteroscedastic noise process and performance is demonstrated numerically. In contrast, this paper rigorously develops the grid-less variational line spectral estimation (VALSE) based approach in heteroscedastic noise environments.

In \cite{VALSE}, VALSE is proposed under homogenous noise, which automatically estimates the number of sources, the nuisance parameters of the prior distribution and noise variance. In addition, in contrast to the previous works which outputs the point estimates of DOAs only, VALSE treats the DOAs as random parameters and outputs the posterior probability density function (PDF) of the DOAs. In \cite{MVALSE}, multi-snapshot VALSE (MVALSE) is developed for Case $\rm \uppercase\expandafter{\romannumeral1}$, and sequential MVALSE (Seq-MVALSE) is also proposed to perform sequential estimation. While in this work, noise assumptions corresponding to Case $\rm \uppercase\expandafter{\romannumeral2}$-$\rm \uppercase\expandafter{\romannumeral4}$ are studied, and three algorithms termed as MVALSE under heteroscedastic noise of snapshots (MVHN-S), MVALSE under heteroscedastic noise of antennas (MVHN-A), MVALSE under heteroscedastic noise of both snapshots and antennas (MVHN) are proposed. Compared to VALSE and MVALSE under homogenous noise, the variants of MVASLE, i.e., the proposed three algorithms, are rederived. In particular, intermediate quantities defined later in (\ref{J-H}) is coupled with the noise variance, whereas intermediate quantities of MVALSE defined in \cite[eq. (21)]{MVALSE} is independent of the noise variance.

This paper studies the DOAs in heteroscedastic noise environment, including noise Case $\rm \uppercase\expandafter{\romannumeral2}$-$\rm \uppercase\expandafter{\romannumeral4}$. Although the noise variance is a nuisance parameter that we are not interested in, estimating the noise variance is beneficial to DOA estimation. In particular, the novelty of this paper is three fold. First, three high resolution low complexity algorithms termed as MVHN-S, MVHN-A and MVHN corresponding to noise Case $\rm \uppercase\expandafter{\romannumeral2}$-$\rm \uppercase\expandafter{\romannumeral4}$ are rigorously developed. It is shown that the three algorithms can be derived in a unified way, and the relationship of these algorithms are revealed, i.e., MVALSE, MVHN-S and MVHN-A can be obtained through MVHN by averaging the noise variances estimates over antennas and snapshots, antennas, snapshots, respectively. Second, by making performance comparison with conventional beamforming (CBF), sparse Bayesian learning and its variants taking noise variation into account \cite{PeterHN} through numerical simulation and a real data set, the advantages of high resolution and low complexity of MVHN-S and MVHN-A are demonstrated. Besides, the proposed method MVHN-A and MVHN-S taking noise variation into account also shows advantages over MVALSE. Thirdly, through numerical experiments an important observation is obtained: the estimated results of MVHN are very biased and MVHN should be avoided even for noise Case $\rm \uppercase\expandafter{\romannumeral4}$. While for Case $\rm \uppercase\expandafter{\romannumeral1}$-$\rm \uppercase\expandafter{\romannumeral3}$, MVHN, MVHN-S, MVHN-A are preferred, respectively.

\emph{Notation}: Let $\mathcal M$ and $\mathcal N$ be the subsets of $\{1,\cdots,M\}$ and $\{1,\cdots,N\}$, and $|\mathcal M|$ denotes its cardinality. For a matrix ${\mathbf A}\in{\mathbb C}^{M\times N}$, let ${\mathbf A}_{{\mathcal M},{\mathcal N}}$ and $[{\mathbf A}]_{{\mathcal M},{\mathcal N}}$ denote the submatrix by choosing the rows and columns indexed by ${\mathcal M}$ and ${\mathcal N}$, respectively. For $M = N$ and ${\mathcal M} = {\mathcal N}$, ${\mathbf A}_{\mathcal M}$ and $[{\mathbf A}]_{\mathcal M}$ denote the submatrix by choosing both rows and columns indexed by ${\mathcal M}$. Let $\mathbf I_L$ denote the identity matrix of dimension $L$. ${(\cdot)}^{*}$, ${(\cdot)}^{\rm T}$ and ${(\cdot)}^{\rm H}$ are the conjugate, transpose and Hermitian transpose operator, respectively. Let $\Re\{\cdot\}$ return the real part. ${\mathcal {CN}}({\mathbf x};{\boldsymbol \mu},{\boldsymbol \Sigma})$ denote the complex normal distribution of ${\mathbf x}$ with mean ${\boldsymbol \mu}$ and covariance ${\boldsymbol \Sigma}$. $\|\cdot\|_0$ denotes the number of nonzero elements.

\section{Problem Setup}\label{ProSet}
Consider a HLA with $M$ antennas uniformly spaced with a half wavelength separation $d=\lambda/2$ and $L$ snapshots are available\footnote{Extending to the incomplete measurement scenario where only a subset $\boldsymbol\Omega=\{(m,l):\Omega_{m,l}=1\}$ of the subset of $\{(m,l),m=1,\cdots,M;l=1,\cdots,L\}$ is available is straightforward. For simplicity, we study the full measurement scenario.}. For the $l$th snapshot, the noisy measurement ${\mathbf y}_l\in{\mathbb C}^{M}$ can be described as
\begin{align}\label{Org_model}
{\mathbf y}_l =\widetilde{\mathbf z}_l +  \widetilde{\mathbf w}_l,\quad l=1,\cdots,L,
\end{align}
where $\{\widetilde{\mathbf z}_l\}_{l=1}^L$ denotes the noiseless signal defined by
\begin{align}\label{defzl}
\widetilde{\mathbf z}_l\triangleq \sum_{k=1}^{K}{{\mathbf a}(\widetilde{\theta}_k)}{\widetilde x}_{k,l},
\end{align}
$\widetilde{\theta}_k \in [-90,90]^{\circ}$ and ${\widetilde x}_{k,l}$ denote the $k$th DOA and the complex weight coefficient, respectively, ${{\mathbf a}({\theta})}$ is the array steering vector defined as
\begin{align}
{\mathbf a({\theta})} &= [1,{\rm e}^{{\rm j}2\pi d \sin{\theta}/\lambda},\cdots,{\rm e}^{{\rm j}2\pi(M-1)d\sin{\theta}/\lambda}]^{\rm T}\notag\\
&=[1,{\rm e}^{{\rm j}\pi \sin{\theta}},\cdots,{\rm e}^{{\rm j}\pi(M-1)\sin{\theta}}]^{\rm T},
\end{align}
$\widetilde{\mathbf w}_l\in{\mathbb C}^{M\times 1}$ is the additive white Gaussian noise independent of the snapshot $l$. Let $\widetilde{w}_{m,l}$ denote the $m$th element of $\widetilde{\mathbf w}_l$ satisfying $\widetilde{w}_{m,l}\sim {\mathcal {CN}}(\widetilde{w}_{m,l};{0},\nu_{m,l})$. For the variance assumptions of four cases described in Section \ref{Intro}, the variances can be mathematically formulated as follows \cite{PeterHN}:
\begin{description}
\item[Case $\rm \uppercase\expandafter{\romannumeral1}$:] \quad Noise variance $\nu\triangleq\nu_{m,l},~\forall ~m,~\forall ~l$ are the same for both different antennas and snapshots.
\item[Case $\rm \uppercase\expandafter{\romannumeral2}$:] \quad Noise variance $\nu_l\triangleq\nu_{m,l},~\forall ~m$ are the same for different antennas, i.e., noise variance depends only on snapshots.
\item[Case $\rm \uppercase\expandafter{\romannumeral3}$:] \quad Noise variance $\nu_m\triangleq\nu_{m,l},~\forall ~l$ are the same for different snapshots, i.e., noise variance depends only on antennas.
\item[Case $\rm \uppercase\expandafter{\romannumeral4}$:] \quad Noise variance $\nu_{m,l}$ is heteroscedastic across both antennas and snapshots, i.e., noise variance depends on both snapshots and antennas.
\end{description}
The goal of this paper is to estimate the number of sources $K$, DOAs $\widetilde{\boldsymbol\theta}$, the complex weight coefficients $\{\widetilde{\mathbf x}_l\}_{l=1}^L$, and the noiseless signal
\begin{align}
\widetilde{\mathbf Z} \triangleq [\widetilde{\mathbf z}_l,\cdots,\widetilde{\mathbf z}_L]\in{\mathbb C}^{M\times L}
\end{align}
for Case $\rm \uppercase\expandafter{\romannumeral2}$-$\rm \uppercase\expandafter{\romannumeral4}$. In the ensuing section, MVHN for Case $\rm \uppercase\expandafter{\romannumeral4}$ is derived and the relationship between MVHN and its variants are then revealed.

\begin{table*}[h!t]
    \begin{center}
\caption{The number of unknown parameters for Case $\rm \uppercase\expandafter{\romannumeral1}$-$\rm \uppercase\expandafter{\romannumeral4}$}\label{Num_para}
\tiny
    \begin{tabular}{|c|c|c|c|c|c|c|}
            \hline
             Scenario    &Case $\rm \uppercase\expandafter{\romannumeral1}$&Case $\rm \uppercase\expandafter{\romannumeral2}$&Case $\rm \uppercase\expandafter{\romannumeral3}$&Case $\rm \uppercase\expandafter{\romannumeral4}$\\ \hline
            $\gamma$ & $K/(2ML)+K/M+{\bf 1/(2ML)}$ & $K/(2ML)+K/M+{\bf 1/(2M)}$ & $K/(2ML)+K/M+{\bf 1/(2L)}$&$K/(2ML)+K/M+{\bf 1/2}$ \\ \hline
             \end{tabular}
    \end{center}
\end{table*}

Before deriving the MVHN, the ratio of the number of unknowns to the number of measurements in real-valued sense defined as $\gamma$ is calculated. Since each measurement is complex-valued, the number of all measurements in real-valued sense is $2ML$. For Case I, the unknown parameters are DOAs ${\boldsymbol \theta}\in {\mathbb R}^K$, complex amplitudes ${\mathbf X}\in {C}^{K\times L}$ and noise variance $\sigma^2$, and the number of unknowns in real-valued sense is $K + 2KL + 1$. Similarly, the numbers of unknowns are $K+2KL + L$, $K + 2KL + M$ and $K + 2KL + ML$ for Cases II-IV, respectively. The results are summarized in Table \ref{Num_para}. In general, the larger the $\gamma$ is, the more likely the algorithm tends to overfit. Thus for Case $\rm \uppercase\expandafter{\romannumeral1}$-$\rm \uppercase\expandafter{\romannumeral4}$, it is expected that the variance of the frequencies estimates of MVHN will be the smallest and the frequencies estimates of MVHN are more likely to be biased, as the noise variances are estimated for each snapshot and antenna, as shown later in Section \ref{Simulation}.

\section{MVHN Algorithm}\label{MVHNS}
In this section, MVHN algorithm for noise Case $\rm \uppercase\expandafter{\romannumeral4}$ is developed. First, the probabilistic formulation similar to \cite{VALSE} is introduced. Then, MVHN is developed.

Before deriving the MVHN algorithm, we reparameterize the model by defining
\begin{align}
\omega=\pi\sin\theta,
\end{align}
where $\omega\in [-\pi,\pi]$ is termed as the frequency. Note that $\omega$ and $\theta$ is a one-to-one correspondence, and $\theta$ can be calculated through $\omega$ as $\theta=\sin^{-1}(\omega/\pi)$, where $\sin^{-1}(\cdot)$ denotes the inverse of $\sin(\cdot)$. In the following, $\omega$ is inferred instead.
\subsection{Probabilistic Formulation}
Since the number of sources $K$ is unknown, an over complete model is imposed, i.e., the number of sources is assumed to be $N$ and $K\leq N\leq M$. As a result, a pseudo observation model for the $l$th snapshot is obtained as
\begin{align}\label{diffnoise-model}
    {\mathbf y}_l  = \sum_{k=1}^{N}{\mathbf a}(\omega_k)x_{k,l} + {\mathbf w}_l = {\mathbf A}({\boldsymbol\omega}){{\mathbf x}_l} + {\mathbf w}_l,\quad l=1,\cdots,L,
\end{align}
where ${\mathbf x}_l$ denote the complex amplitude of the $l$th snapshot and ${\mathbf x}_l = [{x}_{1,l},\cdots,{x}_{N,l}]^{\rm T}\in{\mathbb C}^{N}$, $\mathbf A({{\boldsymbol\omega}})\in{\mathbb C}^{M\times N}$ is
\begin{align}
\mathbf A({{\boldsymbol\omega}}) = [{\mathbf a}({\omega}_1),\cdots,{\mathbf a}({\omega}_N)]\in{\mathbb C}^{M\times N}.
\end{align}
For the notation simplicity, the array model can be described as
\begin{align}
{\mathbf Y} = {\mathbf A}(\boldsymbol\omega){\mathbf X} + {\mathbf W},
\end{align}
where ${\mathbf Y} = [{\mathbf y}_1,\cdots,{\mathbf y}_L]$ is the measurements, ${\mathbf X} = [{\mathbf x}_1,\cdots,{\mathbf x}_L]$ is the complex weight coefficient matrix and ${\mathbf W} = [{\mathbf w}_1,\cdots,{\mathbf w}_L]$ is the noise.

For the $k$th source, let the prior distribution be $p(\omega_k)$. In general, uninformative prior distribution is used and $p(\omega_k)=1/(2\pi)$. For the proposed over complete model, binary hidden variables $\{s_k\}_{k=1}^N\in \{0,1\}^N$ are introduced to promote the sparsity. Specifically, let $s_k=1$ denote the $k$th frequency being active, i.e., the complex weight coefficient ${\mathbf X}_{k,:}$ satisfies $\|{\mathbf X}_{k,:}\|_2\neq0$, otherwise deactive and $\|{\mathbf X}_{k,:}\|_2=0$. Given ${s_k}$, the complex weight coefficient ${\mathbf X}_{k,:}$ is supposed to follow Bernoulli Gaussian distribution
\begin{align}\label{xkdef}
p({\mathbf X_{k,:}}|s_k;\tau) = (1 - s_k){\delta}({\mathbf X}_{k,:}) + s_k{\mathcal {CN}}({\mathbf X}_{k,:};{\mathbf 0},\tau{\mathbf I}_L).
\end{align}
For the prior distribution of $s_k$, Bernoulli distribution is used. Let $\rho$ denote the probability that the $k$th component is active, i.e.,
\begin{align}\label{skdef}
p(s_k;\rho)=\rho^{s_k}(1-\rho)^{1-s_k}.
\end{align}

From measurement model (\ref{diffnoise-model}), the likelihood function $p({\mathbf Y}|{\boldsymbol \omega},{\mathbf X};{\boldsymbol\nu})$ is
\begin{align}\label{llhood}
p({\mathbf Y}|{\boldsymbol \omega},{\mathbf X};{\boldsymbol\nu})=\prod\limits_{l=1}^L {\mathcal {CN}}({\mathbf y}_l;{\mathbf A}({\boldsymbol\omega}){{\mathbf x}_l},\boldsymbol\Sigma_l).
\end{align}
where ${\boldsymbol\nu} = [{\boldsymbol\nu}_1,\cdots,{\boldsymbol\nu}_L]\in{\mathbb C}^{M\times L}$, $\boldsymbol\Sigma_l = {\rm diag}(\boldsymbol\nu_l)$ and  ${\boldsymbol\nu}_l = [{\nu}_{l,1},\cdots,{\nu}_{l,M}]^{\rm T}$.
As a result, the type II ML estimation of the nuisance parameters is
\begin{align}\label{MLunt}
\left(\hat{\rho}_{\rm ML},\hat{\tau}_{\rm ML},\hat{\boldsymbol\nu}_{\rm ML}\right)=\underset{\rho,\tau,{\boldsymbol\nu}}{\operatorname{argmax}}~p({\mathbf Y};\rho,\tau,{\boldsymbol\nu}),
\end{align}
where $p({\mathbf Y};\rho,\tau,{\boldsymbol\nu})$ is the marginalized likelihood function
\begin{align}
&p({\mathbf Y};\rho,\tau,{\boldsymbol\nu})\notag\\
=&\int p({\mathbf Y}|{\boldsymbol \omega},{\mathbf X};{\boldsymbol\nu}) \prod\limits_{k=1}^N\left(p(\omega_k)p({\mathbf X_{k,:}}|s_k;\tau)p(s_k;\rho)\right){\rm d}{\mathbf s}{\rm d}{\mathbf X}{\rm d}{\boldsymbol \omega}.
\end{align}
Given that $\left(\hat{\rho}_{\rm ML},\hat{\tau}_{\rm ML},\hat{\boldsymbol\nu}_{\rm ML}\right)$ are estimated, the maximum a posterior (MAP) estimate is
\begin{align}\label{MLint}
\left(\hat{\boldsymbol \omega}_{\rm MAP},\hat{\mathbf X}_{\rm MAP},\hat{\mathbf s}_{\rm MAP}\right)=\underset{{\boldsymbol \omega},{\mathbf X},{\mathbf s}}{\operatorname{argmax}}~p({\boldsymbol \omega},{\mathbf X},{\mathbf s}|{\mathbf Y};\hat{\rho}_{\rm ML},\hat{\tau}_{\rm ML},\hat{\boldsymbol\nu}_{\rm ML}),
\end{align}
where the posterior PDF $p({\boldsymbol \omega},{\mathbf X},{\mathbf s}|{\mathbf Y};\hat{\rho}_{\rm ML},\hat{\tau}_{\rm ML},\hat{\boldsymbol \nu}_{\rm ML})$ is
\begin{align}
p({\boldsymbol \omega},{\mathbf X},{\mathbf s}|{\mathbf Y};\hat{\rho}_{\rm ML},\hat{\tau}_{\rm ML},\hat{\boldsymbol \nu}_{\rm ML})\propto &p({\mathbf Y}|{\boldsymbol \omega},{\mathbf X};\hat{\boldsymbol \nu}_{\rm ML}) \prod\limits_{k=1}^N\left(p(\omega_k)p({{\mathbf X}_{k,:}}|s_k;\hat{\tau}_{\rm ML})p(s_k;\hat{\rho}_{\rm ML})\right)
\end{align}
Obviously, solving either (\ref{MLunt}) or (\ref{MLint}) is intractable. As a result, a variational Bayesian approach is adopted.

Let $\boldsymbol\Theta=(\omega_1,\cdots,\omega_N,({\mathbf X},{\mathbf s}))$ be the set of all latent variables. To develop the MVHN algorithm, we approximate the posterior PDF $p({\boldsymbol\Theta}|{\mathbf Y})$ as $q({\boldsymbol\Theta}|{\mathbf Y})$ by minimizing the Kullback-Leibler divergence between them defined as ${\rm KL}(q({\boldsymbol\Theta}|{\mathbf Y})||p({\boldsymbol\Theta}|{\mathbf Y}))$ which equals to maximize \cite[pp.~732-733]{Murphy}
\begin{align}\label{Obj-Func}
{\mathcal L}\left(q({\boldsymbol\Theta}|{\mathbf Y})\right) = {\rm E}_{q({\boldsymbol\Theta}|{\mathbf Y})}\left[\ln{\tfrac{p({\mathbf Y},{\boldsymbol\Theta})}{q({\boldsymbol\Theta}|{\mathbf Y})}}\right].
\end{align}
Here $q({\boldsymbol\Theta}|{\mathbf Y})$ is supposed to have the following structure
\begin{align}\label{Appro-Pos-Dis}
q({\boldsymbol\Theta}|{\mathbf Y}) = \prod_{k=1}^Nq(\omega_k|{\mathbf Y})q({\mathbf X}|{\mathbf Y},{\mathbf s})q(\mathbf s),
\end{align}
where $q(\mathbf s)$ is a delta function given by $q(\mathbf s) = \delta({\mathbf s}-{\mathbf s}_0)$, and the joint PDF of ${\mathbf Y}$ and ${\boldsymbol\Theta}$ is
\begin{align}\label{totalequ}
p({\mathbf Y},{\boldsymbol\Theta})=p({\mathbf Y}|{\boldsymbol \omega},{\mathbf X};{\boldsymbol \nu}) \prod\limits_{k=1}^N\left(p(\omega_k)p({\mathbf X_{k,:}}|s_k;{\tau})p(s_k;{\rho})\right).
\end{align}
Maximizing $\mathcal L$ with respect to all the factors is intractable. MVHN iteratively optimizes $\mathcal L$ over each factor $q({\boldsymbol\Theta}_i|{\mathbf Y})$, $i=1,\cdots,N+1$ separately with others being fixed. Thus $q({\boldsymbol\Theta}_i|{\mathbf Y})$ is calculated separately and the posterior approximation $q({\boldsymbol\Theta}_i|{\mathbf Y})$ is calculated as \cite[pp. 735, eq. (21.25)]{Murphy}
\begin{align}\label{upexpression}
\ln q({\boldsymbol\Theta}_i|{\mathbf Y})={\rm E}_{q({{\boldsymbol\Theta}\setminus{\boldsymbol\Theta}_i}|{\mathbf Y})}[\ln p({\mathbf Y},{\boldsymbol\Theta})]+{\rm const},
\end{align}
where the expectation is taken with respect to all the variables ${\boldsymbol\Theta}$ except ${\boldsymbol\Theta}_i$.

Before deriving the MVHN algorithm, some definitions are introduced. ${\widehat\omega}_i$ is defined as the mean direction of $\rm{e}^{{\rm j}\omega_i}$ \cite{Direc} and $\widehat{\mathbf a}_i$ is the estimation of $\mathbf a(\widetilde\omega_i)$ which will be used to give the estimation of weights and reconstructed signals, i.e.,
\begin{subequations}\label{freq-est}
\begin{align}\label{freq-est-a}
    &{\widehat\omega}_i = {\rm arg}\left({\rm E}_{q(\omega_i|{\mathbf Y})}[{\rm e}^{{\rm j}\omega_i}]\right),\\\label{freq-est-b}
    &\widehat{\mathbf a}_i = {\rm E}_{q{(\omega_i|\mathbf Y)}}[\mathbf a(\omega_i)],~i\in\{1,...,N\}.
\end{align}
\end{subequations}
We denote $\widehat{\mathbf A} = [\widehat{\mathbf a}_1,\cdots,\widehat{\mathbf a}_N]$.
The posterior PDF of ${\mathbf X}$ is
\begin{align}\label{posterior-weight}
q({\mathbf X}|{\mathbf Y}) = \int q({\mathbf X},{\mathbf s}|{\mathbf Y})\delta({\mathbf s}-{\mathbf s}_0){\rm d}{\mathbf s} = q({\mathbf X}|{\mathbf Y};{\mathbf s}_0).
\end{align}
Analogously, the mean and covariance of the weights for the $l$th snapshot are estimated as
\begin{subequations}\label{weight-est}
\begin{align}
    &\widehat{{\mathbf x}}_{l} = {\rm E}_{q({\mathbf {X|Y}})}[{\mathbf x}_{l}],\\
    &\widehat{\mathbf C}_l = {\rm E}_{q{(\mathbf {X|Y}})}[{{\mathbf x}_{l}{\mathbf x}_{l}^{\rm H}}] - {\widehat{\mathbf x}}_{l}\widehat{\mathbf x}_{l}^{\rm H},~l=1,...,L,
\end{align}
\end{subequations}
and $\widehat{\mathbf X} = [\widehat{\mathbf x}_1,\cdots,\widehat{\mathbf x}_L]$. Let $\mathcal S$ be the set of the active element indices of $\mathbf s$, i.e.,
\begin{align}
\mathcal S = \{i|1\leq i\leq N,s_i = 1\},
\end{align}
and $\widehat{\mathbf s}$ be the estimate of $\mathbf s$, then the estimated model order is $\widehat{K} = |\widehat{\mathcal S}|$. According to (\ref{diffnoise-model}), the noise-free signal is reconstructed as
\begin{align}
\widehat{\mathbf Z}= \widehat{\mathbf A}_{:,\widehat{\mathcal S}}\widehat{\mathbf X}_{\widehat{\mathcal S},:}.
\end{align}
\subsection{Inferring the Posterior PDF of Frequencies}\label{EstFreq}
In this section, we maximize $\mathcal L$ with respect to the factor $q({\omega_i|\mathbf Y})$ for $i = 1,\cdots,N$. For $i\notin{\mathcal S}$, $q(\omega_i|\mathbf Y)$ is kept unchanged during iteration. According to (\ref{upexpression}), for $i\in{\mathcal S}$, $\ln q(\omega_i|\mathbf Y)$ can be calculated as
\begin{small}
\begin{align}\label{Pos-freq-dis}
&\ln q(\omega_i|\mathbf Y) = {\rm E}_{q({\boldsymbol\Theta}\setminus{\omega_i}|{\mathbf Y})}[\ln p({\mathbf Y}, {\boldsymbol\Theta})] + {\rm const}\notag\\
=&{\rm E}_{q({\boldsymbol\Theta}\setminus{\omega_i}|{\mathbf Y})}\left[\ln(p(\boldsymbol\omega)p(\mathbf s)p(\mathbf {X|s})p(\mathbf {Y}|\boldsymbol \omega,\mathbf X))\right] + {\rm const}\notag\\
=&{\rm E}_{q({\boldsymbol\Theta}\setminus{\omega_i}|{\mathbf Y})}\left[\sum_{j=1}^N \ln p(\omega_j)+\sum_{j=1}^N \ln p(s_j)+\ln p(\mathbf {X|s})\right]
+{\rm E}_{q({\boldsymbol\Theta}\setminus{\omega_i}|{\mathbf Y})}\left[\sum_{l=1}^L\ln p({\mathbf y}_l|\boldsymbol\omega,{\mathbf x}_l;\boldsymbol\Sigma_l)\right] + {\rm const}\notag\\
=&{\rm E}_{q({\boldsymbol\Theta}\setminus{\omega_i}|{\mathbf Y})}\left[\sum_{l=1}^L({\mathbf y}_l-\sum_{k\in{\mathcal S}}{\mathbf a}(\omega_k)x_{k,l})^{\rm H}\boldsymbol\Sigma^{-1}_l({\mathbf y}_l-\sum_{k\in{\mathcal S}}{\mathbf a}(\omega_k)x_{k,l})\right]+\ln p(\omega_i)+{\rm const}\notag\\
=&\ln p(\omega_i)+\Re\left\{{\boldsymbol \eta}_i^{\rm H}\mathbf a(\omega_i)\right\} + {\rm const},
\end{align}
\end{small}
where the complex vector ${\boldsymbol\eta}_i$ is
\begin{align}
{\boldsymbol\eta}_i=\sum\limits_{l=1}^L{\boldsymbol \eta}_{i,l},
\end{align}
and ${\boldsymbol \eta}_{i,l}$ is
\begin{align}\label{yita-i}
{\boldsymbol \eta}_{i,l} = 2\boldsymbol\Sigma^{-1}_l\left[\left({\mathbf y}_l-\sum_{j\neq i}\widehat{\mathbf a}_j{\widehat x}_{j,l}\right){\widehat x}_{i,l}^*-\sum_{j\neq i}[{\widehat{\mathbf C}}_l]_{j,i}\widehat{\mathbf a}_j\right],
\end{align}
which can be viewed as the weighted sum of $2(({\mathbf y}_l-\sum_{j\neq i}\widehat{\mathbf a}_j{\widehat x}_{j,l}){\widehat x}_{i,l}^*-\sum_{j\neq i}[{\widehat{\mathbf C}}_l]_{j,i}\widehat{\mathbf a}_j)$ with respect to the inverse noise variance ${\rm diag}({\boldsymbol\Sigma}^{-1}_l)$.
Given the posterior distribution (\ref{Pos-freq-dis}), (\ref{freq-est}) can not be evaluated in closed form. Thus, by referring to \cite[Heuristic $2$]{VALSE}, $q(\omega_i|\mathbf Y)$ is approximated as a von Mises distribution
\begin{align}\label{approxVM}
q(\omega_i|\mathbf Y)\approx {\mathcal {VM}}(\omega;\widehat{\mu}_i,\widehat{\kappa}_i),
\end{align}
which yields analytical results (\ref{freq-est}). The von Mises distribution ${\mathcal {VM}}(\omega;\mu,\kappa)$ is
\begin{align}
{\mathcal {VM}}(\omega;\mu,\kappa) = \frac{1}{2\pi{I_0}(\kappa)}{\rm e}^{\kappa{\cos(\omega-\mu)}},
\end{align}
where $\mu$ and $\kappa$ are the mean direction and concentration parameters, $I_p(\cdot)$ is the modified Bessel function of the first kind and the order $p$ \cite[p.~348]{Direc}. For a given von Mises distribution,
${\rm arg}({\rm E}_{{\mathcal {VM}}(\omega;\mu,\kappa)}[{\rm e}^{{\rm j}\omega}]) = {\rm arg}\left({\rm e}^{{\rm j}\mu}\frac{I_1(\kappa)}{I_0(\kappa)}\right) = \mu = {\rm E}_{{{\mathcal {VM}}(\omega;\mu,\kappa)}}[\omega]$. In addition, ${\rm E}[{\rm e}^{{\rm j}m\omega}]={\rm e}^{{\rm j}m\mu}(I_m(\kappa)/I_0(\kappa))$ \cite[pp.~26]{Direc}. Therefore, $\widehat{\omega}_i$ and $\widehat{\mathbf a}_i$ (\ref{freq-est}) can be calculated analytically\footnote{An alternative approach is to approximate $q(\omega_i|\mathbf Y)$ as $q(\omega_i|\mathbf Y)\approx \delta(\omega_i-\widehat{\mu}_i)$, which corresponds to the point estimates of the frequencies, and then $\widehat{\mathbf a}_i$ is calculated to be ${\mathbf a}(\widehat{\omega}_i)$. This estimation approach yields the VALSE-pt algorithm \cite{VALSE}. Actually, the larger $\kappa$ of von Mises distribution, the closer $|{\rm E}_{q(\omega|{\mathbf Y})}[{\rm e}^{{\rm j}\omega}]|$ is to one. Conversely, the smaller $\kappa$ is, the closer $|{\rm E}_{q(\omega|{\mathbf Y})}[{\rm e}^{{\rm j}\omega}]|$ is to zero. Hence, for VALSE algorithm, $\widehat{\mathbf a}_i$ reflects the uncertainty of frequency estimation which will influence the model order estimation. For VALSE-pt, it gives full certainty for all frequencies which might overestimate the model order. As shown in \cite{VALSE}, VALSE-pt performs worse than VALSE. Hence, the first approach is adopted in this paper.}.

\subsection{Inferring the Posterior PDF of Weights and Support}\label{EstWeight}
Then $\mathcal L$ is maximized w.r.t. $q({\mathbf X},{\mathbf s}|{\mathbf Y})$.
For $l=1,\cdots,L$, define the matrices $\mathbf J_l$ and $\mathbf H$ as
\begin{subequations}\label{J-H}
\begin{align}
&[{\mathbf J}_l]_{i,j}= 	
\begin{cases}
{\rm tr}(\boldsymbol\Sigma^{-1}_l),&i=j\\
{\widehat{\mathbf a}}^{\rm H}_i{\boldsymbol\Sigma^{-1}_l}{\widehat{\mathbf a}}_j,&i\neq{j}
\end{cases},\quad i,j\in\{1,\cdots,N\},\\
&{\mathbf H}_{:,l} = \widehat{\mathbf A}^{\rm H}{\boldsymbol\Sigma^{-1}_l}{\mathbf y_l},
\end{align}
\end{subequations}
where $[{\mathbf J}_l]_{i,j}$ denotes the $(i,j)$th element of $\mathbf J_l$.

According to (\ref{upexpression}), $q({\mathbf X},{\mathbf s}|{\mathbf Y})$ can be calculated as
\begin{small}
\begin{align}
&\ln q({\mathbf X},{\mathbf s}|{\mathbf Y}) = {\rm E}_{q({\boldsymbol\Theta}\setminus{({\mathbf X,s})}|{\mathbf Y})}\left[
\ln p({\mathbf Y},{\boldsymbol\Theta})\right]+ {\rm const}\notag\\
=& {\rm E}_{q{(\boldsymbol\omega|{\mathbf Y}})}[\sum_{i=1}^N \ln p(s_i) + \ln p(\mathbf {X|s})+\ln p(\mathbf {Y}|\boldsymbol \omega,\mathbf X)]+{\rm const}\notag\\
=&||{\mathbf s}||_0\ln\frac{\rho}{1-\rho} + ||{\mathbf s}||_0L\ln\frac{1}{\pi\tau}- \frac{1}{\tau}\sum_{k\in {\mathcal S}} {\mathbf X}_{{ k},:}{\mathbf X}_{{k},:}^{\rm H}+ {\rm const}\notag\\
-&\sum_{l=1}^L{\rm E}_{q{(\boldsymbol\omega|{\mathbf Y}})}\left[({\mathbf y}_l-\sum_{k\in{\mathcal S}}{\mathbf a}(\omega_k)x_{k,l})^{\rm H}\boldsymbol\Sigma^{-1}_l({\mathbf y}_l-\sum_{k\in{\mathcal S}}{\mathbf a}(\omega_k)x_{k,l})\right] \notag\\
\overset{a}{=}&||{\mathbf s}||_0\ln\frac{\rho}{1-\rho} + ||{\mathbf s}||_0L\ln\frac{1}{\pi\tau}- \frac{1}{\tau}\sum_{l=1}^L {\mathbf X}_{{\mathcal S},l}^{\rm H}{\mathbf X}_{{\mathcal S},l}+\sum_{l=1}^L\left[2\Re\{{\mathbf X}^{\rm H}_{{\mathcal S},l}{\mathbf H}_{{\mathcal S},l}\}-{\mathbf X}^{\rm H}_{{\mathcal S},l}[{\mathbf J_l}]_{{\mathcal S}}{\mathbf X}_{{\mathcal S},l}\right] + {\rm const}\notag\\
=&\sum_{l=1}^L({\mathbf X}_{{\mathcal S},l} - \widehat{\mathbf X}_{{\mathcal S},l})^{\rm H}[\widehat{\mathbf C}_l]_{{\mathcal S}}^{-1}({\mathbf X}_{{\mathcal S},l} - \widehat{\mathbf X}_{{\mathcal S},l}) + {\rm const},\label{Ws_pos}
\end{align}
\end{small}
where $\overset{a}{=}$ utilizes $\sum_{k\in {\mathcal S}} {\mathbf X}_{{ k},:}{\mathbf X}_{{k},:}^{\rm H} = \sum_{l=1}^L {\mathbf X}_{{\mathcal S},l}^{\rm H}{\mathbf X}_{{\mathcal S},l}$ and
\begin{subequations}\label{W-C-1}
\begin{align}
&\widehat{\mathbf X}_{{\mathcal S},l} = [\widehat{\mathbf C}_l]_{{\mathcal S}}{\mathbf H}_{{\mathcal S},l},\\
&[\widehat{\mathbf C}_l]_{{\mathcal S}} = \left([{\mathbf J_l}]_{{\mathcal S}}+\frac{{\mathbf I}_{|{\mathcal S}|}}{\tau}\right)^{-1},~l=1,\cdots,L.
\end{align}
\end{subequations}

According to (\ref{posterior-weight}), to calculate $q({\mathbf X}|{\mathbf Y})$, ${\mathbf s}_0$ has to be given. Plugging the postulated PDF $q({\boldsymbol\Theta}|{\mathbf Y})$ (\ref{Appro-Pos-Dis}) in (\ref{Obj-Func}), one has
\begin{align}\notag
&\ln Z({\mathbf s}_0)\triangleq {\mathcal L}(q({\boldsymbol\Theta}|{\mathbf Y});{\mathbf s}_0)= {\rm E}_{q({\boldsymbol\Theta}|{\mathbf Y})}\left[\ln\tfrac{p({\mathbf Y},{\boldsymbol\Theta};{\mathbf s}_0)}{q({\boldsymbol\Theta}|{\mathbf Y};{\mathbf s}_0)}\right]\notag\\
=& {\rm E}_{q({\boldsymbol\Theta}|{\mathbf Y})}\left[\ln p({\mathbf s}) + \ln p({\mathbf X}|{\mathbf s})\right]+{\rm E}_{q({\boldsymbol\Theta}|{\mathbf Y})}\left[\ln p({\mathbf Y}|\boldsymbol\omega,{\mathbf X})- \ln q({\mathbf X}|{\mathbf Y})\right]+{\rm const}\notag\\
=&||{\mathbf s}_0||_0\ln\frac{\rho}{1-\rho}-\sum_{l=1}^L[\ln\det([{\mathbf J}_l]_{{\mathcal S}_0}+\frac{1}{\tau}\mathbf I_{|{\mathcal S}_0|})+||{\mathbf s}_0||_0\ln\frac{1}{\tau}+({\mathbf H}_{{\mathcal S}_0,l}^{\rm H}([{\mathbf J}_l]_{{\mathcal S}_0}+\frac{1}{\tau}{\mathbf I}_{|{\mathcal S}_0|})^{-1}{\mathbf H}_{{\mathcal S}_0,l})]
+{\rm const},\label{lnZ}
\end{align}
where ${\mathcal S}_0$ is the set of the active element indices of ${\mathbf s}_0$.
Thus ${\mathbf s}_0$ should be chosen to maximize $\ln Z({\mathbf s}_0)$ (\ref{lnZ}), i.e.,
\begin{align}
\widehat{\mathbf s}_0 = \underset{{\mathbf s}_0}{\operatorname{argmax}}~\ln Z({\mathbf s}_0).
\end{align}

A naive approach to solve the above problem is to enumerate all the possible binary values of ${\mathbf s}_0$, which costs $O(2^N)$ and is impractical for typical values of $N$. To reduce the computation complexity, a greedy iterative search strategy is proposed to find a local optimum. Given ${\mathbf s}_0$, the strategy proceeds as follows: For each $k=1,\cdots,N$, calculate $\Delta_k=\ln Z({\mathbf s}_0^k)-\ln Z({\mathbf s}_0)$, where ${\mathbf s}_0^k$ is the same as ${\mathbf s}_0$ except that the $k$th element of ${\mathbf s}_0$ is flipped. Let $k^*=\underset{k}{\operatorname {argmax}}~\Delta_k$. If $\Delta_{k^*}>0$, we update ${\mathbf s}_0$ with the $k^*$th element flipped, and ${\mathbf s}_0$ is updated, otherwise $\widehat{\mathbf s}_0$ is obtained as ${\mathbf s}_0$, and the algorithm is terminated. In fact, $\Delta_k$ can be easily calculated and the details are provided in Appendix \ref{modelSelection}. Numerically, the computational complexity of the greedy approach is about $O(\widehat K)$, where $\widehat{K} = |\widehat{\mathcal S}|$ denotes the estimated model order.
\subsection{Estimating the Model Parameters}\label{EstModel}
After updating the frequencies and weights, the model parameters $\boldsymbol\beta = \{{\boldsymbol \nu},~\rho,~\tau\}$ are estimated via maximizing the lower bound ${\mathcal L}(q({\boldsymbol\Theta}|{\mathbf Y});{\boldsymbol \beta})$. Plugging the postulated PDF (\ref{Appro-Pos-Dis}) in (\ref{Obj-Func}), ${\mathcal L}(q{({\boldsymbol\Theta}|{\mathbf Y})};\boldsymbol \beta)$ is
\begin{small}
\begin{align}\label{Para-Lower}
&{\mathcal L}(q{({\boldsymbol\Theta}|{\mathbf Y})};\boldsymbol \beta)= {\rm E}_{q{({\boldsymbol\Theta}|{\mathbf Y}})}\left[\ln{\tfrac{p({\mathbf {Y,}{\boldsymbol\Theta};\boldsymbol \beta})}{q({\boldsymbol\Theta}|{\mathbf Y})}}\right]\notag\\
=& {\rm E}_{q({\boldsymbol\Theta}|{\mathbf Y})}\left[\ln p({\mathbf s}) + \ln p({\mathbf X}|{\mathbf s})+\ln p(\mathbf {Y}|\boldsymbol\omega,\mathbf X)\right]+{\rm const}\notag\\
=&||{\widehat{\mathbf s}||_0\ln\rho} -||{\widehat{\mathbf s}||_0\ln(1-\rho)} + ||\widehat{\mathbf s}||_0L\ln\frac{1}{\pi\tau}-{\rm E}_{q({\mathbf {X|Y}})}\left[\frac{1}{\tau}{\rm tr}(\mathbf X_{{\widehat{\mathcal S}},:}{\mathbf X}^{\rm H}_{\widehat{\mathcal S},:})\right]-\sum_{l=1}^L{\rm E}_{q{(\mathbf {X|Y}})}\left[{\mathbf X}^{\rm H}_{\widehat{\mathcal S},l}[{\mathbf J}_l]_{\widehat{\mathcal S}}{\mathbf X}_{\widehat{\mathcal S},l}\right] \notag\\
+&\sum_{l=1}^L\left(\ln\frac{1}{\det(\pi{\boldsymbol\Sigma_l})}-{\mathbf y}_l^{\rm H}{\boldsymbol\Sigma^{-1}_l}{\mathbf y}_l+2\Re\{(\widehat{\mathbf X}^{\rm H}_{\widehat{\mathcal S},l}{\mathbf H}_{\widehat{\mathcal S},l})\}\right)+ {\rm const}.
\end{align}
\end{small}
Substituting ${\rm E}_{q{(\mathbf {X|Y}})}[{\rm tr}({\mathbf X}_{\widehat{\mathcal S},:}{\mathbf X}^{\rm H}_{\widehat{\mathcal S},:})]={\rm tr}(\widehat{\mathbf X}^{\rm H}_{\widehat{\mathcal S},:}\widehat{\mathbf X}_{\widehat{\mathcal S},:})+\sum_{l=1}^L{\rm tr}([\widehat{\mathbf C}_l]_{\widehat{\mathcal S}})$ and ${\rm E}_{q{(\mathbf {X|Y}})}[{\mathbf X}^{\rm H}_{\widehat{\mathcal S},l}[{\mathbf J}_l]_{\widehat{\mathcal S}}{\mathbf X}_{\widehat{\mathcal S},l}]={\rm tr}\left([{\mathbf J}_l]_{\widehat{\mathcal S}}(\widehat{\mathbf X}_{\widehat{\mathcal S},l}\widehat{\mathbf X}^{\rm H}_{\widehat{\mathcal S},l}+[\widehat{\mathbf C}_l]_{\widehat{\mathcal S}})\right)$ in (\ref{Para-Lower}), one has
\begin{small}
\begin{align}
{\mathcal L}(q{({\boldsymbol\Phi}|{\mathbf Y})};\boldsymbol \beta)=&\sum_{l=1}^L\left(\ln\frac{1}{\det(\pi{\boldsymbol\Sigma_l})}-{\mathbf y}_l^{\rm H}{\boldsymbol\Sigma^{-1}_l}{\mathbf y}_l+2\Re\{(\widehat{\mathbf X}^{\rm H}_{\widehat{\mathcal S},l}{\mathbf H}_{\widehat{\mathcal S},l})\}\right)+N\ln(1-\rho)\notag\\
-&\sum_{l=1}^L{\rm tr}\left([{\mathbf J}_l]_{\widehat{\mathcal S}}(\widehat{\mathbf X}_{\widehat{\mathcal S},l}\widehat{\mathbf X}^{\rm H}_{\widehat{\mathcal S},l}+[\widehat{\mathbf C}_l]_{\widehat{\mathcal S}})\right) ||\widehat{\mathbf s}||_0\left(\ln\frac{\rho}{1-\rho}-L{\rm ln}\tau\right)\notag\\
-&\frac{1}{\tau}\left[{\rm tr}(\widehat{\mathbf X}_{\widehat{\mathcal S},:}\widehat{\mathbf X}^{\rm H}_{\widehat{\mathcal S},:})+\sum_{l=1}^L{\rm tr}([\widehat{\mathbf C}_l]_{\widehat{\mathcal S}})\right]+{\rm const},\label{Lvaluemodpa}
\end{align}
\end{small}
Setting $\frac{\partial\mathcal L}{\partial\rho}=0$ and $\frac{\partial\mathcal L}{\partial\tau}=0$, we have
\begin{align}\label{rou-tau-hat}
\widehat{\rho} =\frac{||\widehat{\mathbf s}||_0}{N},\quad \quad \widehat{\tau} = \frac{{\rm tr}\left({\mathbf X}_{\widehat{\mathcal S},:}^{\rm H}{\mathbf X}_{\widehat{\mathcal S},:}\right)+\sum_{l=1}^L{\rm tr}([\widehat{\mathbf C}_l]_{\widehat{\mathcal S}})}{L||\widehat{\mathbf s}||_0}.
\end{align}
Setting $\frac{\partial\mathcal L}{\partial\nu_{m,l}}=0,~m=1,\cdots,M,~l=1,\cdots,L$, we obtain
\begin{align}\label{nu-MVHN}
\widehat{\nu}_{m,l}& = {|y_{m,l}-{\widehat{\mathbf A}}_{i,\widehat{\mathcal S}}{\widehat{\mathbf X}}_{\widehat{\mathcal S},l}|^2}+{\widehat{\mathbf A}}_{i,\widehat{\mathcal S}}[\widehat{\mathbf C}_l]_{\widehat{\mathcal S}}{\widehat{\mathbf A}}^{\rm H}_{i,\widehat{\mathcal S}}+\sum_{i\in\widehat{\mathcal S}}|\widehat{x}_{i,l}|^2(1-|{\widehat{\mathbf A}}_{m,i}|^2).
\end{align}
It is worth noting that $\widehat{\nu}_{m,l}$ consists of three terms, where the first term is the fitting (residue) error, the second term is the error coming from the complex weight ${\mathbf W}$, and the last term from the frequencies ${\boldsymbol \omega}$. Given that the fitting is perfect, the weight estimate or the frequency is estimated exactly ($\widehat{\mathbf C}_l\rightarrow {\mathbf 0}$ or $\widehat{\kappa}_i \rightarrow \infty$ (\ref{approxVM})), the corresponding three terms will diminish.

The initializations of MVHN are the same as \cite{MVALSE} and MVHN is outlined in Algorithm \ref{MVHN}.
\begin{algorithm}[ht]
\caption{Outline of MVHN algorithm}\label{MVHN}
\textbf{Input:}~~Signal matrix $\mathbf Y$\\
\textbf{Output:}~~The model order estimate $\widehat{K}$, frequencies estimate $\widehat{\boldsymbol\omega}_{\widehat{\mathcal S}}$, complex weights estimate $\widehat{\mathbf X}_{\widehat{\mathcal S},:}$ and reconstructed signal $\widehat{\mathbf Z}$
\begin{algorithmic}[1]
\STATE Initialize $\widehat{\boldsymbol\nu},\widehat\rho,\widehat\tau$~and~$q(\omega_i|{\mathbf Y}),i\in\{1,\cdots,N\}$
\STATE \textbf{repeat}
\STATE ~~~~Calculate $q({\mathbf X},{\mathbf s}|{\mathbf Y})$ and update~$\widehat{\mathbf s},\widehat{\mathbf X}_{\widehat{\mathcal S},:}$ (Section \ref{EstWeight})
\STATE ~~~~Update the parameters $\widehat\rho$, $\widehat\tau$ (\ref{rou-tau-hat}) and $\widehat{\nu}_{l,m},~l=1,\cdots,L,~m=1,\cdots,M$ (\ref{nu-MVHN})
\STATE ~~~~Calculate $q({\boldsymbol\omega}|{\mathbf Y})$ and update $\widehat{\boldsymbol\omega}$ (Section \ref{EstFreq})
\STATE \textbf{until} stopping criterion
\STATE \textbf{return} $\widehat{K}$, $\widehat{\boldsymbol\omega}_{\widehat{\mathcal S}}$, $\widehat{\mathbf X}_{\widehat{\mathcal S},:}$ and $\widehat{\mathbf Z}$
\end{algorithmic}
\end{algorithm}

\section{The Variant Algorithms of MVHN}
Now we focus on noise Case $\rm \uppercase\expandafter{\romannumeral2}$-$\rm \uppercase\expandafter{\romannumeral3}$, which are special cases of Case $\rm \uppercase\expandafter{\romannumeral4}$. The variants of MVHN algorithm, termed as MVHN-S and MVHN-A are derived and relationship between MVALSE, MVHN-S, MVHN-A and MVHN are revealed.

The only one minor difference between MVHN and other three algorithms is the estimation of noise variances in (\ref{nu-MVHN}).
For MVALSE of Case $\rm \uppercase\expandafter{\romannumeral1}$ in \cite{MVALSE}, we have ${\mathbf J}_1 = \cdots = {\mathbf J}_L = {\mathbf J}$ and $\widehat{\mathbf C}_1 = \cdots = \widehat{\mathbf C}_L = \widehat{\mathbf C}$. The estimate of noise variance is
\begin{align}\label{nu-MVALSE}
\widehat{\nu} =& {||\mathbf Y-{\widehat{\mathbf A}}_{:,\widehat{\mathcal S}}{\widehat{\mathbf X}}_{\widehat{\mathcal S},:}||^2_{\rm F}}/({ML})+{{\rm tr}(\mathbf J_{\widehat{\mathcal S}}\widehat{{\mathbf C}}_{\widehat{\mathcal S}})}/{M}+\sum_{i\in\widehat{\mathcal S}}\sum_{l=1}^L|\widehat{x}_{i,l}|^2(1-{||\widehat{\mathbf a}_i||_2^2}/M)/L,
\end{align}
which is consistent with \cite{MVALSE}. For Case $\rm \uppercase\expandafter{\romannumeral2}$, we have $\nu_{1,l} = \cdots = \nu_{M,l} = \nu_l$. The estimates of noise variances are
\begin{align}\label{nu-MVHN-S}
\widehat{\nu}_l = &\frac{1}{M}\sum_{m=1}^M\left({|y_{m,l}-{\widehat{\mathbf A}}_{i,\widehat{\mathcal S}}{\widehat{\mathbf X}}_{\widehat{\mathcal S},l}|^2}+{\widehat{\mathbf A}}_{i,\widehat{\mathcal S}}[\widehat{\mathbf C}_l]_{\widehat{\mathcal S}}{\widehat{\mathbf A}}^{\rm H}_{i,\widehat{\mathcal S}}\right)+\frac{1}{M}\sum_{m=1}^M\sum_{i\in\widehat{\mathcal S}}|\widehat{x}_{i,l}|^2(1-|{\widehat{\mathbf A}}_{m,i}|^2).
\end{align}
For Case $\rm \uppercase\expandafter{\romannumeral3}$, we have $\nu_{m,l} = \cdots = \nu_{m,L} = \nu_m$. The estimates of noise variances are
\begin{align}\label{nu-MVHN-A}
\widehat{\nu}_m = &\frac{1}{L}\sum_{l=1}^L\left({|y_{m,l}-{\widehat{\mathbf A}}_{i,\widehat{\mathcal S}}{\widehat{\mathbf X}}_{\widehat{\mathcal S},l}|^2}+{\widehat{\mathbf A}}_{i,\widehat{\mathcal S}}[\widehat{\mathbf C}_l]_{\widehat{\mathcal S}}{\widehat{\mathbf A}}^{\rm H}_{i,\widehat{\mathcal S}}\right)+ \frac{1}{L}\sum_{l=1}^L\sum_{i\in\widehat{\mathcal S}}|\widehat{x}_{i,l}|^2(1-|{\widehat{\mathbf A}}_{m,i}|^2).
\end{align}
According to (\ref{nu-MVALSE}), (\ref{nu-MVHN-S}), (\ref{nu-MVHN-A}) and (\ref{nu-MVHN}), the variance estimates under Case $\rm \uppercase\expandafter{\romannumeral1}$-$\rm \uppercase\expandafter{\romannumeral3}$ can be derived from Case $\rm \uppercase\expandafter{\romannumeral4}$, i.e.,
\begin{subequations}
\begin{align}
&\widehat\nu = \sum_{m=1}^M\sum_{l=1}^{L}\widehat{\nu}_{m,l}/(ML),\label{nu-Case1}\\
&\widehat{\nu}_l = \sum_{m=1}^M\widehat{\nu}_{m,l}/M,\label{nu-Case2}\\
&\widehat{\nu}_m = \sum_{l=1}^L\widehat{\nu}_{m,l}/L,\label{nu-Case3}
\end{align}
\end{subequations}
Thus, the estimate of noise variance for Case $\rm \uppercase\expandafter{\romannumeral1}$ is the average of Case $\rm \uppercase\expandafter{\romannumeral4}$ on both antennas and snapshots from (\ref{nu-Case1}). In addition, the noise variance estimates of Case $\rm \uppercase\expandafter{\romannumeral2}$ and Case $\rm \uppercase\expandafter{\romannumeral3}$ are the average of Case $\rm \uppercase\expandafter{\romannumeral4}$ on antennas and snapshots from (\ref{nu-Case2}) and (\ref{nu-Case3}), respectively.
To sum up, the three algorithms MVALSE, MVHN-S and MVHN-A can be viewed as the variants of MVHN.

The computational complexity of MVHN and its variants are analyzed. For the MVHN shown in Algorithm \ref{MVHN}, the computation complexity is dominated by both the estimation of model order in Sec. \ref{EstWeight} and the approximation of $q(\omega_i|{\mathbf Y}),i\in{\widehat{\mathcal S}}$ as von Mises distribution. According to \cite{VALSE, MVALSE}, the whole complexity is $O(NL{\widehat K}^2+MNL{\widehat K})$ for each iteration. Compared to the variants of MVHN,  MVHN differs only in noise estimation, thus the whole complexities of the variants of MVHN are still $O(NL{\widehat K}^2+MNL{\widehat K})$. It is also worth noting that the proposed methods involve the computation of the ratio of two modified Bessel functions of the first kind shown in (\ref{freq-est-b}), and the complexity can be reduced by looking up tables. For the CBF approach, the computation complexity is $O(MLN_g)$, where $N_g$ denotes the number of grids and  usually satisfies $N_g>M$. For the SBL based approaches, the computation complexity is dominated by the linear minimum mean squared error estimation which involves a matrix inversion. For the SBL and SBLHN-A, only a single matrix inversion shared among all the snapshots is needed for each iteration, and the computation complexity is $O(N_g^3)$. While for SBLHN and SBLHN-S, one needs to calculate the matrix inversions for each snapshot, and the computation complexity is $O(LN_g^3)$. The results are also summarized in Table \ref{comp_comp}. It can be seen that the computation complexity of CBF is the lowest, followed by MVHN, SBL and SBLHN-A, SBLHN and SBLHN-S.

\begin{table*}[h!t]
    \begin{center}
\caption{The computational complexity of algorithms with $T$ being the number of iterations}\label{comp_comp}
        \begin{tabular}{|c|c|c|c|c|}
          \hline
          Algorithms & CBF & MVHN and its variants & SBL and SBLHN-A & SBLHN and SBLHN-S \\ \hline
          Computation complexity & $O(MLN_g)$ & $O((NL{\widehat K}^2+MNL{\widehat K})T)$ & $O(N_g^3T)$ & $O(LN_g^3T)$ \\
          \hline
        \end{tabular}
    \end{center}
\end{table*}

To provide the benchmark performance of the four algorithms for noise Cases I-IV, the CRB is derived and is postponed to Appendix \ref{DerivationCRB}.

\section{Numerical Simulation}\label{Simulation}
In this section, four numerical experiments are conducted to evaluate the performance of the four algorithms, i.e., MVALSE, MVHN-S, MVHN-A and MVHN algorithms. For the first simulation, the high resolutions of the MVHN-S and MVHN-A are illustrated, compared to the low-resolution CBF approach. For the second simulation, better reconstruction performance is demonstrated, compared to the SBL based approach. For the third and fourth simulations, the performances of all the four algorithms versus the nominal SNR or the fluctuation strength of the noise variance (both are defined in the next paragraph) are investigated. For convenience, the variants of SBL for case $\rm \uppercase\expandafter{\romannumeral2}$, case $\rm \uppercase\expandafter{\romannumeral3}$ and case $\rm \uppercase\expandafter{\romannumeral4}$ are named as SBLHN-S, SBLHN-A and SBLHN, respectively.

\emph{Simulation Setup}: The magnitudes and phases of the complex weight coefficients are generated i.i.d. from normal distribution $\mathcal N(1,0.2)$ and uniform distribution $\mathcal U(-\pi,\pi)$, respectively. For each numerical simulation, the frequencies $\{\widetilde\omega\}_{i=1}^K$ and complex amplitudes $\{\widetilde{\mathbf x}_l\}_{l=1}^L$ are fixed for all the Monte Carlo (MC) trials. We define nominal signal-to-noise ratio (SNR) as ${\rm SNR} \triangleq 10{\rm log}(||\widetilde{\mathbf Z}||_{\rm F}^2/(\widetilde\nu_0ML)$, where $\widetilde\nu_0$ is the nominal noise variance, $||\cdot||_{\rm F}$ is the Frobenius norm and we denote $\widetilde\nu_{0,{\rm dB}}=10\log\widetilde\nu_0$. Note that for Case I, the definitions of the SNR (nominal SNR) and noise variance are precise, while for Case II-IV, either the SNR (nominal SNR) or noise variance is inaccurate for Case II-IV. To describe Case I-IV quantitatively, we introduce the terms nominal  SNR and noise variance. The noise variance of four cases are generated as follows:
\begin{description}
\item[Case $\rm \uppercase\expandafter{\romannumeral1}$:] \quad Noise variance $\widetilde\nu$ equals to nominal noise variance, i.e., $\widetilde\nu = \widetilde\nu_0$.
\item[Case $\rm \uppercase\expandafter{\romannumeral2}$:] \quad $10\log\widetilde\nu_l$ is generated from uniform distribution ${\mathcal U}(\widetilde\nu_{0,{\rm dB}},\widetilde\nu_{0,{\rm dB}}+\Delta_{\nu})$, $l=1,\cdots,L$.
\item[Case $\rm \uppercase\expandafter{\romannumeral3}$:] \quad $10\log\widetilde\nu_m$ is generated from uniform distribution ${\mathcal U}(\widetilde\nu_{0,{\rm dB}},\widetilde\nu_{0,{\rm dB}}+\Delta_{\nu})$, $m=1,\cdots,M$.
\item[Case $\rm \uppercase\expandafter{\romannumeral4}$:] \quad $10\log\widetilde\nu_{m,l}$ is generated from uniform distribution ${\mathcal U}(\widetilde\nu_{0,{\rm dB}},\widetilde\nu_{0,{\rm dB}}+\Delta_{\nu})$, $m=1,\cdots,M,~l=1,\cdots,L$,
\end{description}
where $\Delta_{\nu}$ is the strength of noise fluctuation in dB which characterizes the fluctuation of noise variances. The Algorithm \ref{MVHN} stops when  $||\widehat{\mathbf X}^{(t-1)} - \widehat{\mathbf X}^{(t)}||_2/||\widehat{\mathbf X}^{(t-1)}||_2 < 10^{-6}$ or $t > 500$, where $t$ is the index of the iteration.

\subsection{Variants of MVHN versus CBF}\label{VSCBF}
\begin{figure*}
  \centering
  \subfigure[]{
    \label{Case1PDFa}
    \includegraphics[width=160mm]{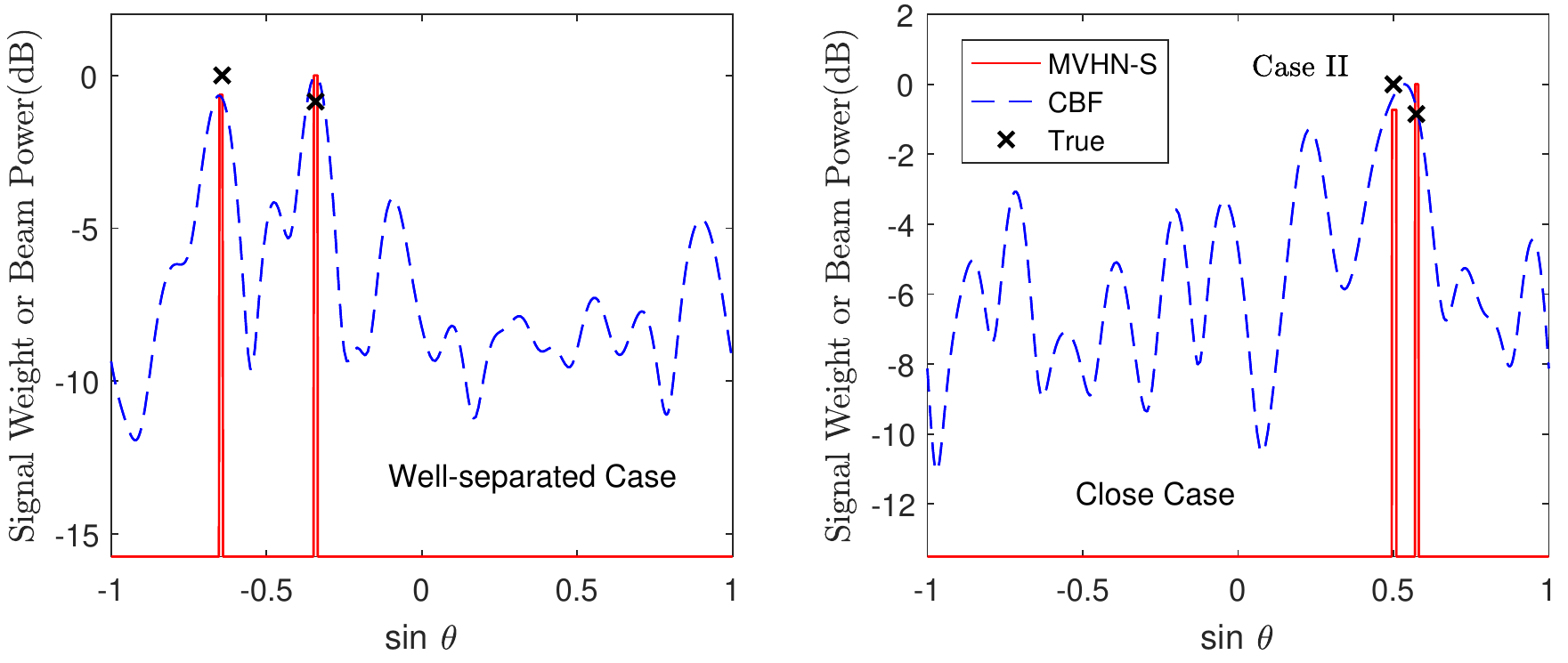}}
  \subfigure[]{
    \label{Case1PDFb}
    \includegraphics[width=160mm]{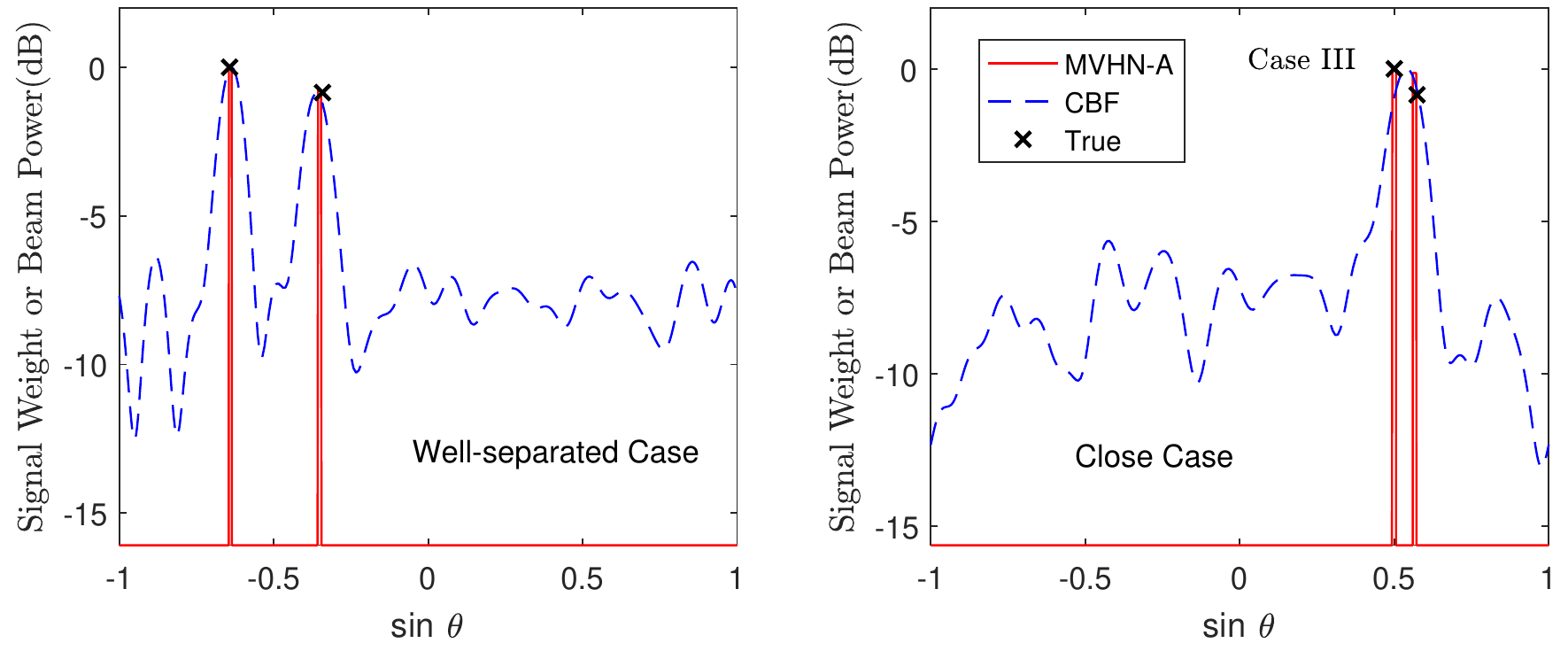}}
  \caption{The performance of CBF, MVHN-S and MVHN-A. Left, the two frequencies are well separated. Right, the two frequencies are close. (a) Signals are generated in Case $\rm \uppercase\expandafter{\romannumeral2}$. Both the weight and beam power are normalized for better illustration in this figure. (b) Signals are generated in Case $\rm \uppercase\expandafter{\romannumeral3}$.}
    \label{Case1PDF} 
\end{figure*}
The performances of the MVHN-S and MVHN-A are evaluated under Case $\rm \uppercase\expandafter{\romannumeral2}$ and $\rm \uppercase\expandafter{\romannumeral3}$. Parameters are set as follows: $M=N=20$, $L=10$, $K=2$, SNR $=5$ dB and $\Delta_{\nu}=15$ dB. The true DOAs are $\widetilde{\boldsymbol \theta}=[-40^{\circ};-20^{\circ}]^{\rm T}$ and  $\widetilde{\boldsymbol \theta}=[30^{\circ};35^{\circ}]^{\rm T}$ for well-separated case and close case, respectively. The number of grids for CBF is set as $361$, i.e., the grid spacing is $0.5^{\circ}$. For the MVHN-S and MVHN-A, the mean estimates plus the $\pm3$ standard deviations through the posterior PDF are plotted\footnote{For a von Mises distribution ${\mathcal {VM}}(\theta,\mu,\kappa)$ with $\kappa>10$, the von Mises distribution approaches a normal distribution ${\mathcal N}(\theta,\mu,1/\kappa)$ with high accuracy \cite{Direc}. In our setting, the concentration parameter $\kappa$ of the posterior PDF output by the algorithms usually satisfies $\kappa>10^2$. Thus the mean estimate plus the $\pm3$ standard deviation can be approximated as $\mu\pm 3/\sqrt{\kappa}$.}, which can be viewed as bins with the height being the estimated signal weight in the plot, where the signal weight in dB corresponding to the $k$th target is defined as $20\log\left(\|\widehat{\mathbf X}_{k,:}\|_2/\sqrt{L}\right)$. For better comparison, we normalize signal weight curve and beam power. The results are presented in Fig. \ref{Case1PDF}. Note that for both Case $\rm \uppercase\expandafter{\romannumeral2}$ and $\rm \uppercase\expandafter{\romannumeral3}$, CBF performs well under well-separated scenario, and can not resolve the two DOAs under close scenario. In contrast, MVHN-S and MVHN-A perform well under both well-separated and close scenarios, demonstrating the high resolution of MVHN-S and MVHN-A. In addition, the running time averaged over $10$ MC trials on a desktop computer with an Intel(R) Core(TM) i5-8265U 1.60GHz CPU of the algorithms are evaluated, and results are shown in Table \ref{time_CBF}. It can be seen that the running time of CBF is lower than that of MVHN-S and MVHN-A.
\begin{table}[h]
\centering
\caption{Execution time of CBF, MVHN-S and MVHN-A in seconds.}\label{time_CBF}
\begin{tabular}{|c|c|c|c|}
\hline
\diagbox [width=5em,trim=l] {Scenario}{Alg.}&CBF & MVHN-S& MVHN-A\\\hline
Case $\rm \uppercase\expandafter{\romannumeral2}$ (well-separated case) & $0.0021$ & $1.20$ &---\\\hline
Case $\rm \uppercase\expandafter{\romannumeral2}$ (close case) & $0.0016$ & $3.36$ &---\\\hline
Case $\rm \uppercase\expandafter{\romannumeral3}$ (well-separated case) & $0.0018$ & ---  &$1.48$\\\hline
Case $\rm \uppercase\expandafter{\romannumeral3}$ (close case)  & $0.0021$ & ---  &$3.33$\\\hline
\end{tabular}
\end{table}
\subsection{Variants of MVHN versus variants of SBLHN}\label{VSSBL}
In this section, the performances of the MVHN and SBL and their variants are evaluated under four Case $\rm \uppercase\expandafter{\romannumeral2}$-$\rm \uppercase\expandafter{\romannumeral4}$. It is worth noting that SBL based approaches are on-grid and the number of grids $N_g$ is set as $N_g=361$, i.e., the same as the setting of CBF and the grid spacing is $0.5^{\circ}$. Since SBL based approaches do not output the number of targets exactly, the top $K$ peaks are regarded as the targets.

\begin{figure*}
  \centering
  \subfigure[]{
    \label{Case1PDF2a}
    \includegraphics[width=170mm]{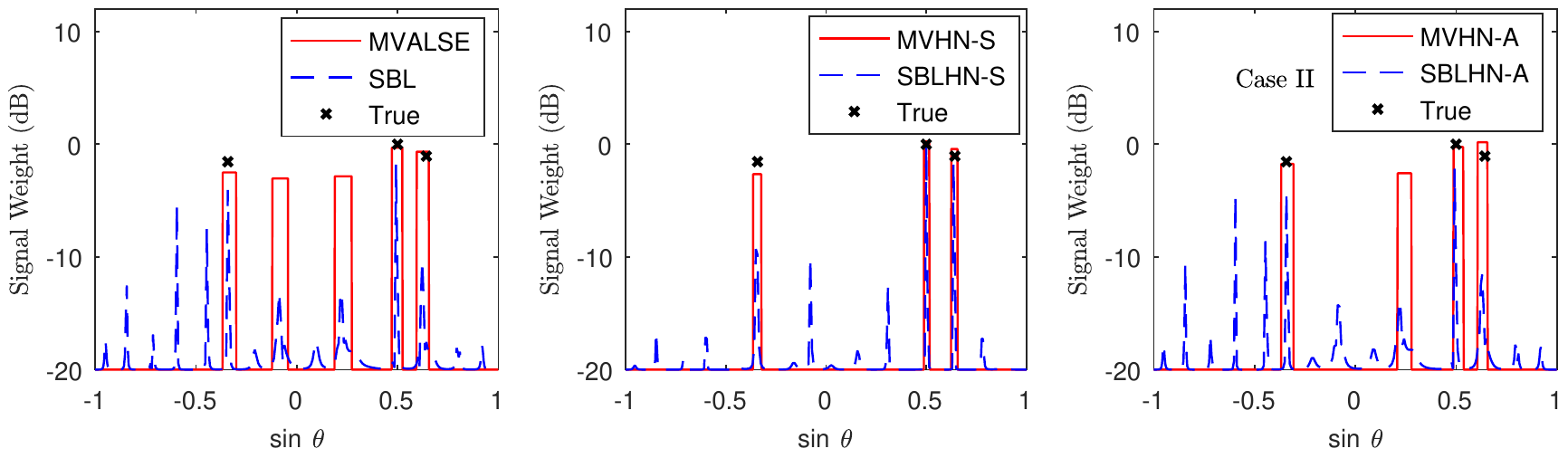}}
  \subfigure[]{
    \label{Case1PDF2b}
    \includegraphics[width=170mm]{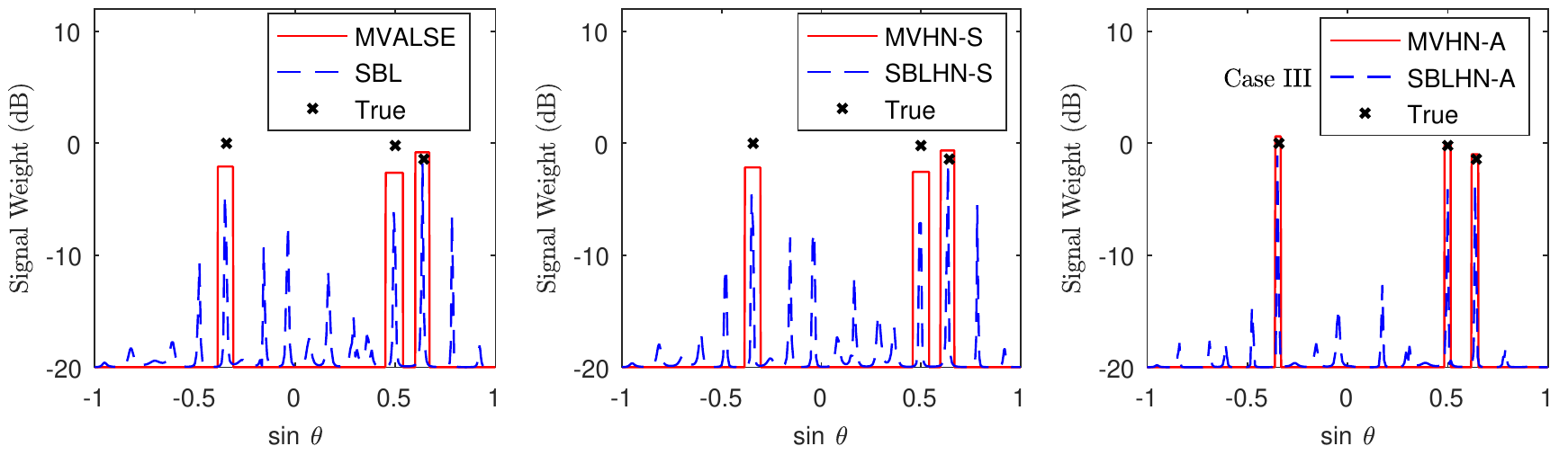}}
  \subfigure[]{
    \label{Case1PDF2c}
    \includegraphics[width=170mm]{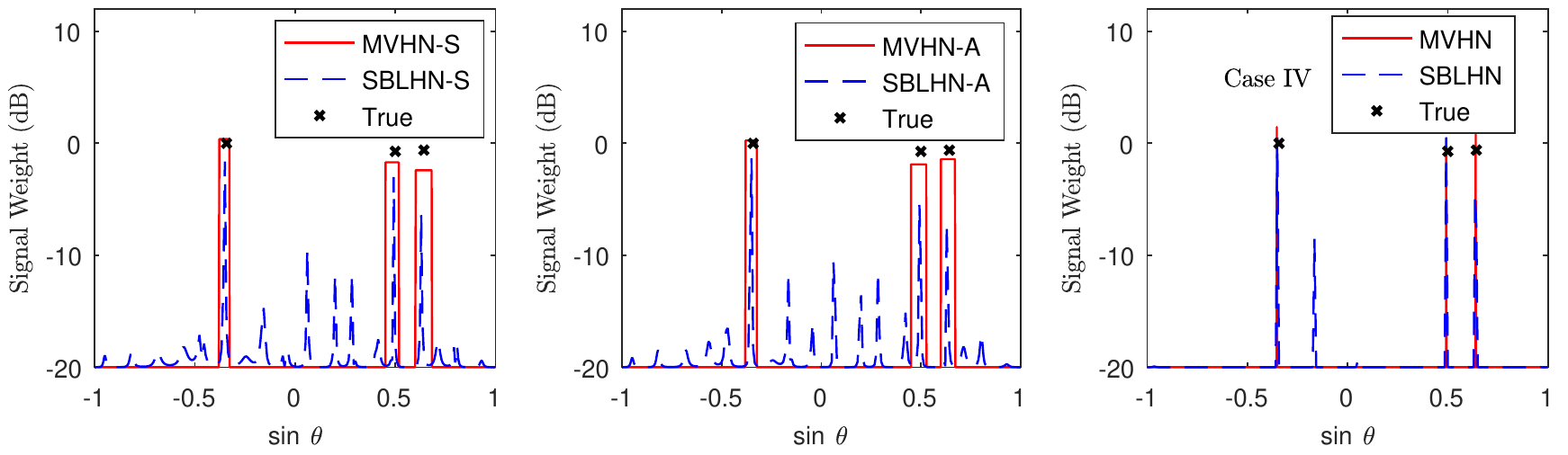}}
  \caption{The performance of SBL and MVHN and its variants in a single simulation. (a) Signals are generated in Case $\rm \uppercase\expandafter{\romannumeral2}$. (b) Signals are generated in Case $\rm \uppercase\expandafter{\romannumeral3}$. (b) Signals are generated in Case $\rm \uppercase\expandafter{\romannumeral4}$.}
    \label{Case1PDF2} 
\end{figure*}

Parameters are set as follows: $M=N=20$, $L=10$,  $K=3$, SNR $=5$ dB, $\Delta_{\nu}=15$ dB. The true DOAs are $\widetilde{\boldsymbol \theta}=[-20^{\circ};30^{\circ};40^{\circ}]^{\rm T}$. Note that the first and the second targets are well separated, while the second and third targets are close.   The results are presented in Fig. \ref{Case1PDF2}.

For Case $\rm \uppercase\expandafter{\romannumeral2}$, the results are shown in Fig. \ref{Case1PDF2a}. The numbers of targets output by MVALSE and MVHN-A are $5$ and $4$, respectively, which yield $2$ and $1$ false targets. In detail, MVALSE and MVHN-A estimate all the true targets. While MVHN-S successfully estimates the model order and recovers the DOAs in high accuracy. In addition, the standard deviations characterized by the width of the bins output by the MVHN-S are the smallest, and the three bins cover the true targets. For the SBL, SBLHN-S and SBLHN-A, they output many spurious components and the top $K$ peaks are $[-0.59,-0.34,0.49]^{\rm T}$, $[-0.35,0.5,0.64]^{\rm T}$, $[-0.59,-0.34,0.49]^{\rm T}$, and the true $\sin \widetilde{\boldsymbol \theta}$ is $\sin \widetilde{\boldsymbol \theta}=[-0.34;0.5;0.64]^{\rm T}$. Thus both SBL and SBLHN-A miss the third target corresponding to DOA $40^{\circ}$, while mistaken that there exists a target with DOA $\arcsin(-0.59)\approx-36^{\circ}$. In summary, MVHN-S performs best, followed by SBLHN-S.

For Case $\rm \uppercase\expandafter{\romannumeral3}$, the results are shown in Fig. \ref{Case1PDF2b}. MVALSE, MVHN-S and MVHN-A successfully estimate the model order and recover the DOAs in high accuracy. In addition, the standard deviations output by the MVHN-A is the smallest, followed by MVHN-S and MVALSE, and the three bins cover the true targets. For the SBL, SBLHN-S and SBLHN-A, they output many spurious components and the top $K=3$ peaks are $[-0.35,0.49,0.64]^{\rm T}$, $[-0.35,0.64,0.78]^{\rm T}$, $[-0.35,0.5,0.64]^{\rm T}$, and the true $\sin \widetilde{\boldsymbol \theta}$ is $\sin \widetilde{\boldsymbol \theta}=[-0.34;0.5;0.64]^{\rm T}$. Thus SBL and SBLHN-A recover the true DOAs, while SBLHN-S misses the second target corresponding to DOA $30^{\circ}$, while mistaken that there exists a target with DOA $\arcsin(0.78)\approx 51^{\circ}$. It can be seen that SBLHN-A performs better than SBL, as the fourth peak output by SBLHN-A is $8.5$ dB lower than the third peak, while the fourth peak output by SBL is $0.36$ dB lower than the third peak. The signal weight estimation accuracy of MVHN-A is higher than that of SBL-A, as SBL diffuses the whole energy into the other false targets. In summary, MVHN-A performs best in Case $\rm \uppercase\expandafter{\romannumeral3}$.

For Case $\rm \uppercase\expandafter{\romannumeral4}$, the results are shown in Fig. \ref{Case1PDF2c}. MVHN-S, MVHN-A and MVHN successfully estimate the model order and recover the DOAs in high accuracy. In addition, the standard deviations output by the MVHN is the smallest, followed by MVHN-A and MVHN-S.  However, the three bins output by MVHN do not cover the true targets, while the three bins output by MVHN-S and MVHN-A cover the true targets. This shows that the estimates output by MVHN is very biased due to the absence of the averaging of the noise, which will be further validated in Fig. \ref{Case2SNR}. For the SBL, SBLHN-S and SBLHN-A, the top $K=3$ peaks are $[-0.35,0.5,0.64]^{\rm T}$, $[-0.35,0.5,0.64]^{\rm T}$, $[-0.35,0.5,0.64]^{\rm T}$, and the true $\sin \widetilde{\boldsymbol \theta}$ is $\sin \widetilde{\boldsymbol \theta}=[-0.34;0.5;0.64]^{\rm T}$. Thus SBL, SBLHN-S and SBLHN-A recover the true DOAs. In detail, it can be seen that SBLHN performs better than SBLHN-S and SBLHN-A, as the fourth peak output by SBLHN is $3.9$ dB lower than the third peak, while the fourth peaks output by SBLHN-S and SBLHN-A are $3.1$ dB and $3.3$ dB lower than the third peak. It can be seen that the signal weight accuracy of MVHN is higher than that of SBLHN, as SBLHN also diffuses the whole energy into the other false targets. In summary, MVHN-S, MVHN-A and MVHN work in Case $\rm \uppercase\expandafter{\romannumeral3}$ and the estimates of MVHN are biased.

From Fig. \ref{Case1PDF2}, MVHN-S, MVHN-A and MVHN output the narrowest width of the bins for Case $\rm \uppercase\expandafter{\romannumeral2}$, Case $\rm \uppercase\expandafter{\romannumeral3}$ and Case $\rm \uppercase\expandafter{\romannumeral4}$, respectively. In addition, the bins output by MVHN-S and MVHN-A cover the true targets, while the bins output by MVHN do not cover the true targets. The above numerical results lead to the following observations: For the noise Case $\rm \uppercase\expandafter{\romannumeral2}$, MVHN-S is preferred, for the noise Case $\rm \uppercase\expandafter{\romannumeral3}$, MVHN-A is preferred, and for the noise Case $\rm \uppercase\expandafter{\romannumeral4}$, MVHN-S or MVHN-A is preferred and MVHN should be avoided as it tends to overfit and output biased results. More numerical experiments are conducted to corroborate this point in the following subsections.

The average running time of MVALSE, SBL and their variants on the same desktop computer is also evaluated and results are shown in Table \ref{time_SBL}. The running time of MVALSE, MVHN-S and MVHN-A are similar and shorter than that of MVHN. It is found that MVHN usually needs more iterations to converge. As for the SBL and its variants, the running time of SBL and SBLHN-A is similar, while the running time of SBLHN-S and SBLHN is much more longer due to the $L$ times matrix inversion, as analyzed in Table \ref{comp_comp}. To sum up, the running time of MVHN and its variants is shorter than that of SBL and its variants.

\begin{table}[h]
\centering
\caption{Execution times of SBL, MVALSE and their variants averaged over $10$ MC trials in seconds.}\label{time_SBL}
\begin{tabular}{|c|c|c|c|c|c|c|c|c|}
  \hline
  \diagbox [width=5em,trim=l] {Scenario}{Alg.} & SBL & SBLHN-S & SBLHN-A& SBLHN & MVALSE & MVHN-S & MVHN-A & MVHN  \\  \hline
  Case $\rm \uppercase\expandafter{\romannumeral2}$  & 39.72 & 130.13 & 45.30 &--- & 2.69 & 2.17 & 3.71 & --- \\  \hline
  Case $\rm \uppercase\expandafter{\romannumeral3}$ & 38.44 & 124.51 &  43.69 & --- &1.76 & 1.80 & 3.12 & --- \\  \hline
  Case $\rm \uppercase\expandafter{\romannumeral4}$ & --- & 118.71 & 37.31& 167.66 & ---&2.22& 2.35 & 18.21 \\
  \hline
\end{tabular}
\end{table}

\subsection{Performance Versus the Nominal SNR} \label{SimuSNR}
\begin{figure*}
  \centering
  \subfigure[]{
    \label{Case2SNRgen1}
    \includegraphics[width=165mm]{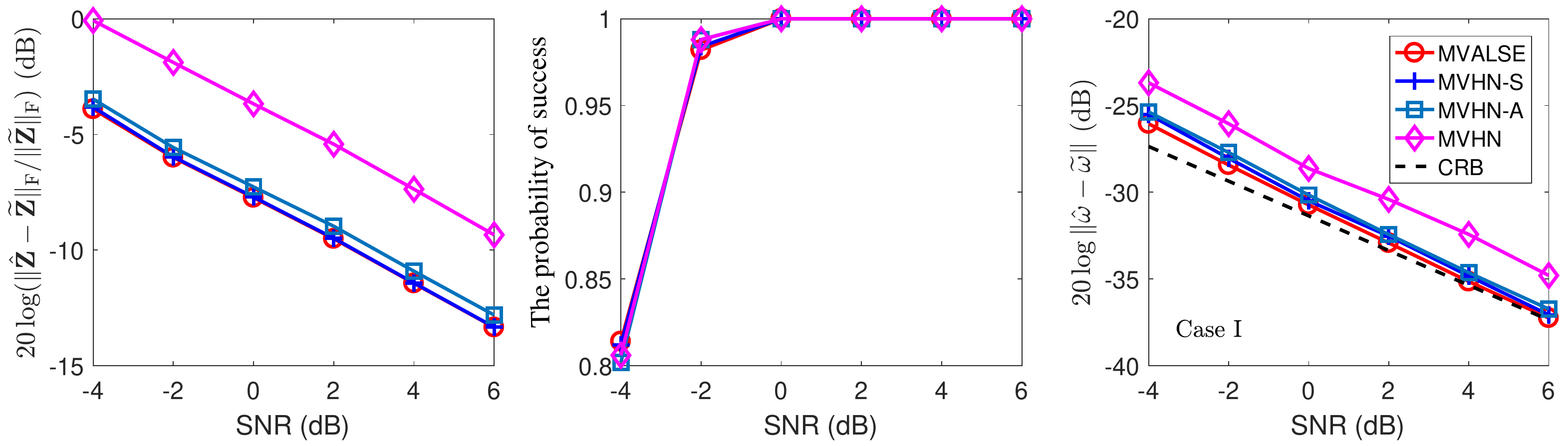}}
  \subfigure[]{
    \label{Case2SNRgen2}
    \includegraphics[width=165mm]{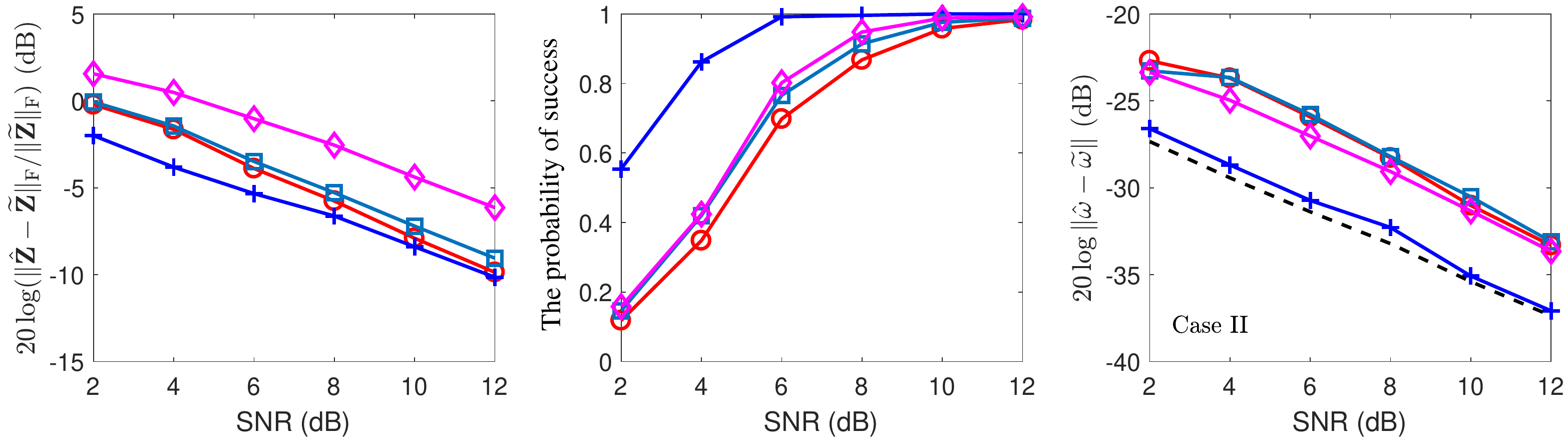}}
  \subfigure[]{
    \label{Case2SNRgen3}
    \includegraphics[width=165mm]{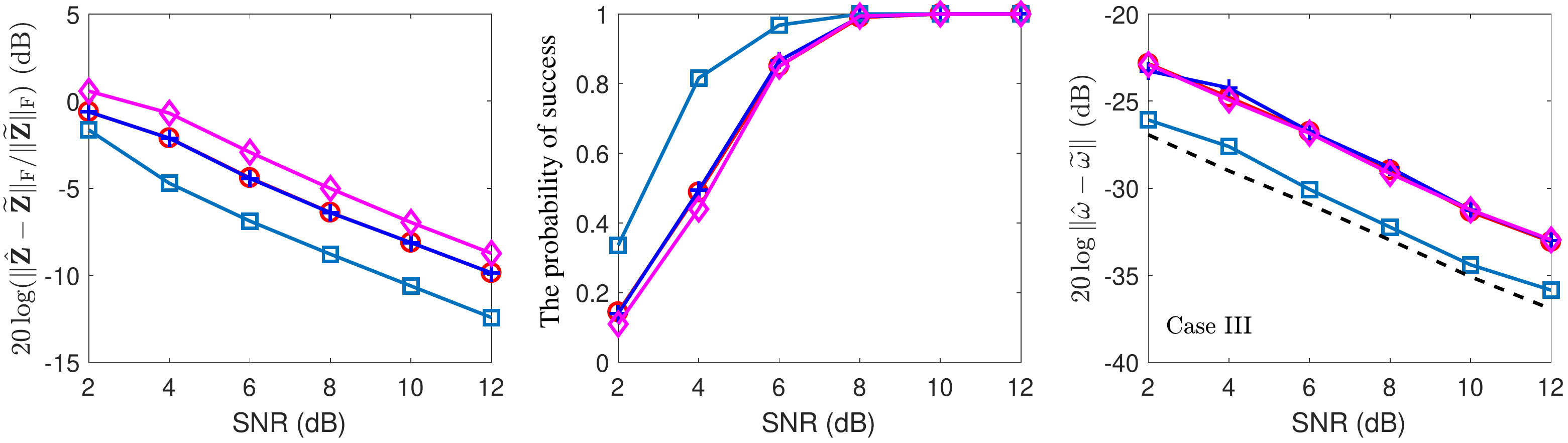}}
  \subfigure[]{
    \label{Case2SNRgen4}
    \includegraphics[width=165mm]{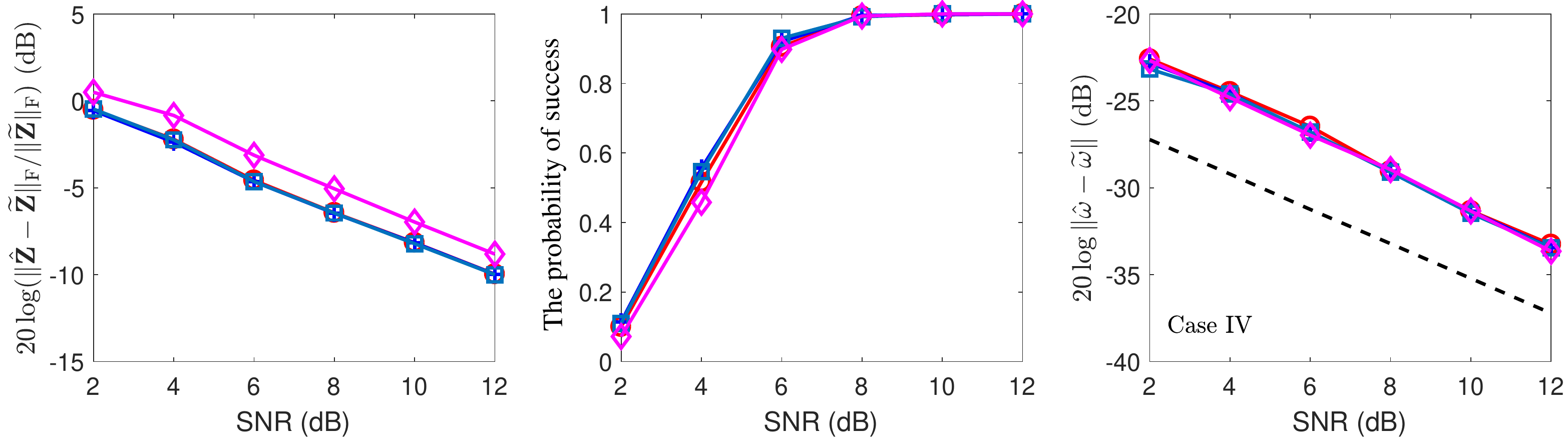}}
 \caption{Performance versus SNR for $M=20$, $L=10$. Fig. \ref{Case2SNRgen1}, \ref{Case2SNRgen2}, \ref{Case2SNRgen3}, \ref{Case2SNRgen4} correspond to noise Case $\rm \uppercase\expandafter{\romannumeral1}$-$\rm \uppercase\expandafter{\romannumeral4}$, respectively.}
  \label{Case2SNR} 
\end{figure*}

In this section, the performances of the four algorithms (MVHN, MVHN-S, MVHN-A, MVALSE) under Case $\rm \uppercase\expandafter{\romannumeral1}$-$\rm \uppercase\expandafter{\romannumeral4}$ with varied nominal SNR are investigated. The normalized mean squared error (NMSE) of $\widehat{\mathbf Z}$ is ${\rm NMSE}(\widehat{\mathbf Z}) \triangleq 10{\rm log}(||\widehat{\mathbf Z}-\widetilde{\mathbf Z}||_{\rm F}^2/||\widetilde{\mathbf Z}||_{\rm F}^2)$ and MSE of $\widehat{\boldsymbol\omega}$ is ${\rm MSE}(\widehat{\boldsymbol\omega}) \triangleq 10{\rm log}(||\widehat{\boldsymbol\omega} - \widetilde{\boldsymbol\omega}||_2^2)$, the correct model order estimated probability ${\rm P}(\widehat{K}=K)$ are adopted as the performance metrics. The frequency estimation error is averaged over the trials in which ${\widehat K} = K$ and $|\widehat\omega_k-\widetilde\omega_k| \leq \pi/N,~\forall k$ for a given simulation point. Parameters are set as follows: $M=N=20$, $L=10$, $K=3$ and the true frequencies are $\widetilde{\boldsymbol \omega}=[-0.1;0.5;2.1]^{\rm T}$. The strength of noise fluctuation is $\Delta_{\nu} = 15$ dB for Case II-IV. While for Case I, the noise variance is fixed and thus $\Delta_{\nu} = 0$ dB. All the results are averaged over $500$ MC trials and results are shown in Fig. \ref{Case2SNR}.

From Fig. \ref{Case2SNRgen1}, MVALSE, MVHN-S and MVHN-A achieve almost the same performances in terms of the signal reconstruction error, model order estimation probability and frequency estimation error, and asymptotically approach the CRB, while MVHN shows some performance degradation under Case $\rm \uppercase\expandafter{\romannumeral1}$. The reason is that the noise variance estimate step of MVHN does not average over the samples and its frequency estimates are very biased, and the amplitude estimates $\widehat{\mathbf X}$ is not accurate, which lead to large frequency estimation error and signal reconstruction error.

For Case $\rm \uppercase\expandafter{\romannumeral2}$, Fig. \ref{Case2SNRgen2} shows that MVHN-S performs best in terms of signal reconstruction error, followed by MVALSE, MVHN-A and MVHN.  For the model order estimation probability, MVHN-S achieves the highest, followed by MVHN, MVHN-A and MVALSE. For the frequency estimation errors, MVHN-S performs best, followed by MVHN, MVHN-A and MVALSE. Besides, the MVHN-S approaches to the CRB as SNR increases. To sum up, MVHN-S performs best in terms of the three criteria, and MVHN performs worst in terms of signal reconstruction error.

For Case $\rm \uppercase\expandafter{\romannumeral3}$, Fig. \ref{Case2SNRgen3} shows that MVHN-A performs best in terms of signal reconstruction error. The performance of MVALSE is comparable to that of MVHN-S, and is better than that of MVHN.  For the model order estimation probability, MVHN-A achieves the highest. The performance of MVALSE is comparable to that of MVHN-S, and is slightly better than that of MVHN. For the frequency estimation errors, MVHN-A performs best, followed by MVHN-S, MVHN, MVALSE. There exists a performance gap between MVHN-A and the CRB. To sum up, MVHN-A performs best in terms of the three criteria, and MVHN still performs worst in terms of signal reconstruction error.

For Case $\rm \uppercase\expandafter{\romannumeral4}$ in Fig. \ref{Case2SNRgen4}, it shows that MVHN-A, MVHN-S and MVALSE perform similarly in terms of signal reconstruction error, and are better than MVHN. For the model order estimation probability, MVHN-A, MVHN-S and MVALSE perform similarly and are slightly better than MVHN. For the frequency estimation errors, the performances of the four algorithms are comparable, and there exists an obvious performance gap compared to the CRB. To sum up, MVHN-A MVHN-S and MVALSE perform better than MVHN in this Case.

From Fig. \ref{Case2SNR}, it is concluded that for Case $\rm \uppercase\expandafter{\romannumeral1}$-$\rm \uppercase\expandafter{\romannumeral3}$, MVALSE, MVHN-S and MVHN-A are preferred, respectively. While for Case $\rm \uppercase\expandafter{\romannumeral4}$, it is still recommended to use MVALSE, MVHN-S, or MVHN-A instead of MVHN.

\subsection{Performance Versus $\Delta_{\nu}$}
\begin{figure*}
  \centering
  \subfigure[]{
    \label{Case2deltagen2}
    \includegraphics[width=170mm]{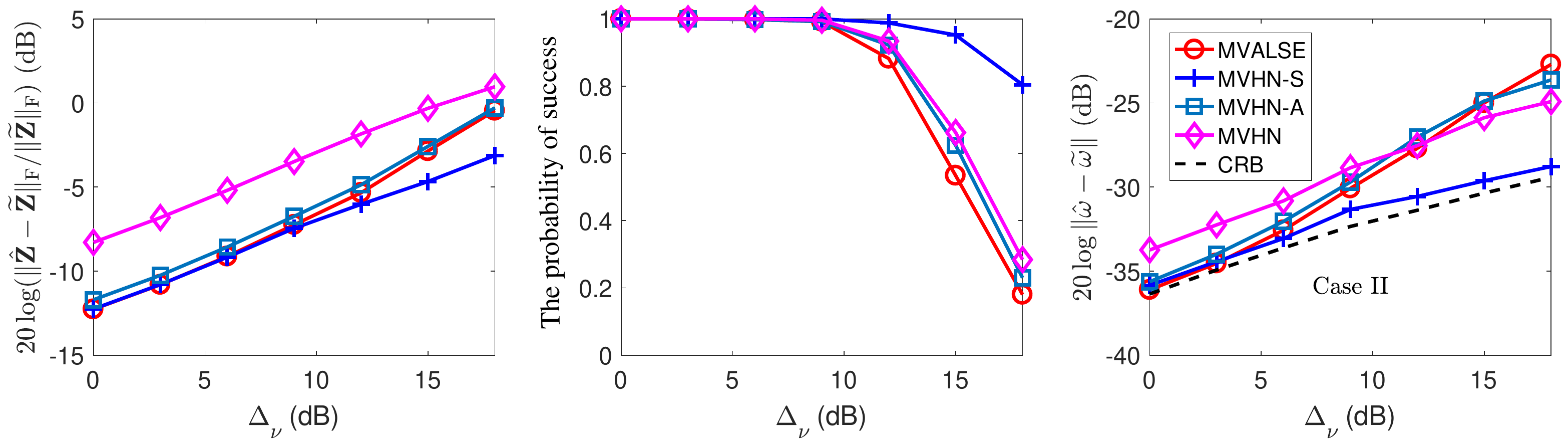}}
  \subfigure[]{
    \label{Case2deltagen3}
    \includegraphics[width=170mm]{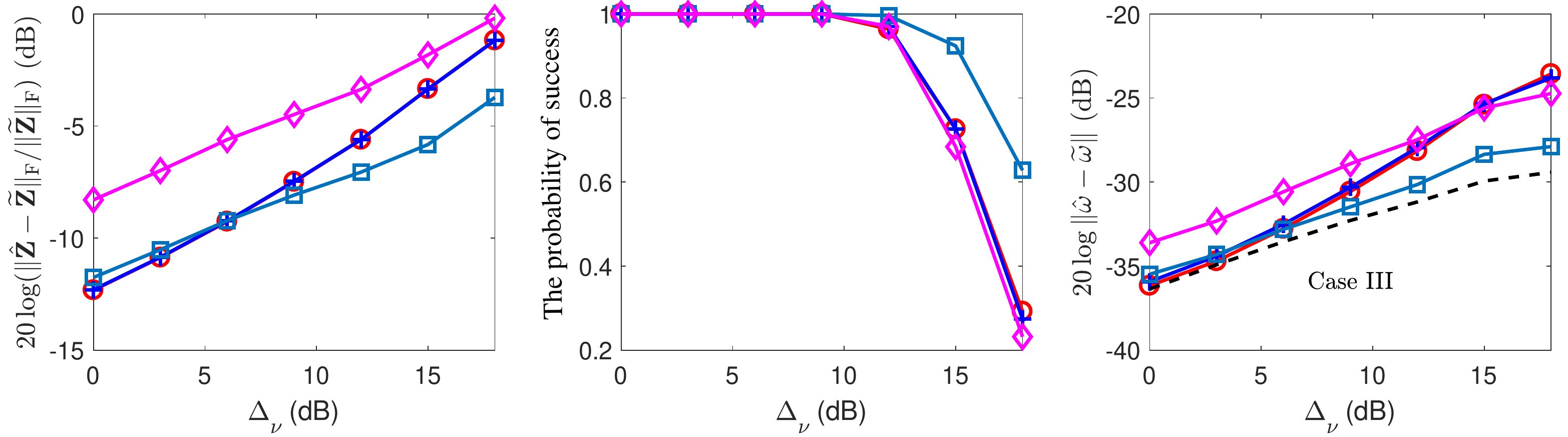}}
  \subfigure[]{
    \label{Case2deltagen4} 
    \includegraphics[width=170mm]{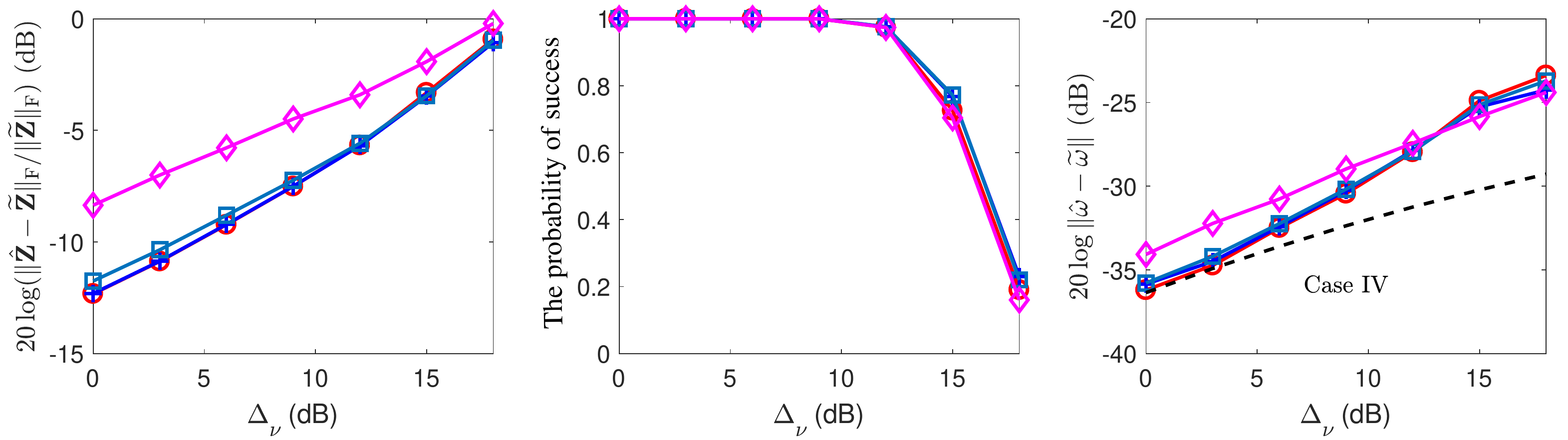}}
 \caption{Performance of algorithms by varying $\Delta_{\nu}$ for Case $\rm \uppercase\expandafter{\romannumeral2}$-$\rm \uppercase\expandafter{\romannumeral4}$ corresponding to Fig. \ref{Case2deltagen2}, \ref{Case2deltagen3}, \ref{Case2deltagen4}. The number of measurements is $M=20$ and the number of snapshots is $L=10$. }
  \label{Case2delta} 
\end{figure*}

The performances of the four algorithms under Case $\rm \uppercase\expandafter{\romannumeral2}$-$\rm \uppercase\expandafter{\romannumeral4}$ are investigated with varied strength of noise fluctuation $\Delta_{\nu}$. Parameters are set as follows: $M=N=20$, $L=10$, $K=3$ and $\widetilde{\boldsymbol \omega}=[-0.1;0.5;2.1]^{\rm T}$. The nominal SNR is ${\rm SNR}=5$ dB and $500$ MC trials are performed. Results are shown in Fig. \ref{Case2delta}.

From Fig. \ref{Case2deltagen2} under Case $\rm \uppercase\expandafter{\romannumeral2}$,  MVHN-S performs best in terms of the signal reconstruction error, followed by MVALSE, MVHN-A and MVHN. In addition, the signal reconstruction performance gap between MVHN-A and MVHN is obvious, which is due to the overfitting of MVHN. MVHN-S achieves the highest model order estimation probability followed by MVHN, MVHN-A and MVALSE. For the frequency estimation error, all the algorithms begin to deviate away from CRB as $\Delta_{\nu}$ increases, especially for the MVALSE, MVHN-A and MVHN. MVHN-S is more robust than other algorithms in terms of the three criteria. In conclusion, the performances of all algorithms degrade as $\Delta_{\nu}$ increases and MVHN-S performs best in Case $\rm \uppercase\expandafter{\romannumeral2}$.

From Fig. \ref{Case2deltagen3} under Case $\rm \uppercase\expandafter{\romannumeral3}$, the three algorithms MVHN-A, MVHN-S and MVALSE perform similar in terms of signal reconstruction error when $\Delta_{\nu}<6$ dB. As $\Delta_{\nu}$ increases, MVHN-A performs better than the other three algorithms. Similar to Fig. \ref{Case2deltagen2}, MVHN performs worst in the whole considered range. For model order estimation probability, MVHN-A performs best, and the remaining algorithms perform similarly. As $\Delta_{\nu}$ increases, the frequency estimation error of all the algorithms degrades and deviates away from CRB. MVHN-A performs best in terms of frequency estimation error. MVHN is more robust than MVALSE and MVHN-S as it degrades slower than those algorithms in terms of signal reconstruction error and frequency estimation error. To sum up, the performances of all algorithms degrade as $\Delta_{\nu}$ increases and MVHN-A performs best in Case $\rm \uppercase\expandafter{\romannumeral3}$.

From Fig. \ref{Case2deltagen4} under Case $\rm \uppercase\expandafter{\romannumeral4}$,
MVALSE, MVHN-S and MVHN-A achieve almost the same performances in terms of the signal reconstruction error, while MVHN performs worst. For model order estimation probability,  all the algorithms have a similar performance. MVALSE, MVHN-S and MVHN-A also perform similarly in terms of frequency estimation error, and begin to deviate away from CRB as $\Delta_{\nu}$ increases. MVHN performs worst when $\Delta_{\nu}<15$ dB and there is always a gap between CRB and its performance. To sum up, even in Case $\rm \uppercase\expandafter{\romannumeral4}$, it is still recommended to use MVALSE, MVHN-S or MVHN-A instead of MVHN.

From Fig. \ref{Case2delta}, it is concluded that MVALSE, MVHN-S and MVHN-A are preferred for Case $\rm \uppercase\expandafter{\romannumeral1}$-$\rm \uppercase\expandafter{\romannumeral3}$, respectively. While for Case $\rm \uppercase\expandafter{\romannumeral4}$, one should prefer MVALSE, MVHN-S or MVHN-A rather than MVHN due to its biasness.

\section{Application: Real Data Analysis}
In this section, the performances of CBF, SBLHN-A, MVALSE and MVHN-A are evaluated with real experimental acoustic data collected at sea. The data were collected by the horizontal linear arrays (HLA) deployed on the seafloor \cite{TCYang}. The array had $32$ elements uniformly spaced with an aperture of $15$ m (design frequency $50$ Hz). A low-frequency J$-15-3$ acoustic source towed by the R/V Endeavor is at around $\sim 12$ m depth and broadcasts continuous tones at $50$ Hz. The ship (with the towed source) was moving outward from the HLA at a direction near broadside of the array. The $10$ snapshots are arranged as a batch and the data generated from the first $16$ elements of the array are chosen. We use CBF, SBLHN-A, MVALSE for performance comparison.

\begin{figure*}
  \centering
  \subfigure[]{
    \label{RealData1}
    \includegraphics[width=80mm]{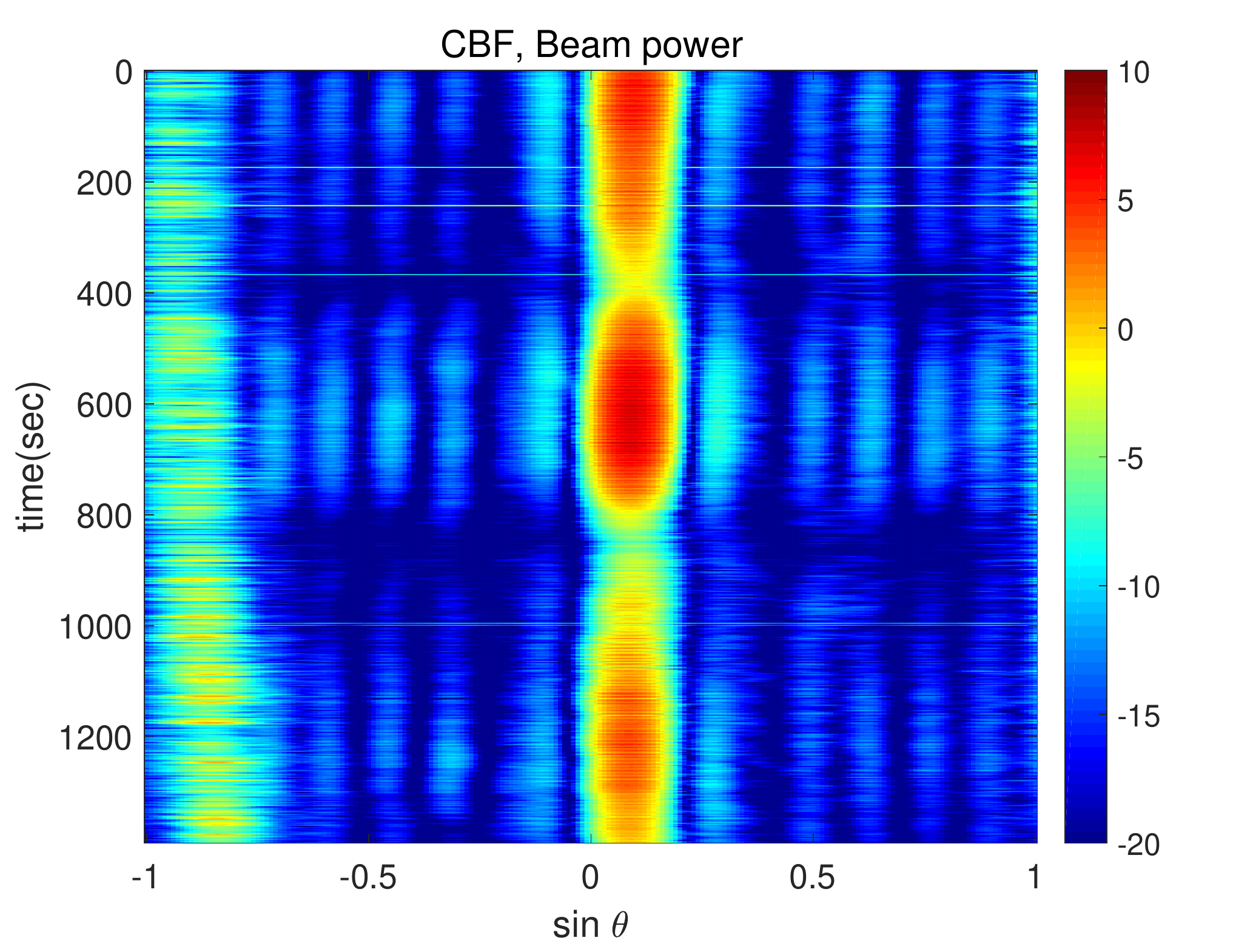}}
  \subfigure[]{
    \label{RealData2}
    \includegraphics[width=80mm]{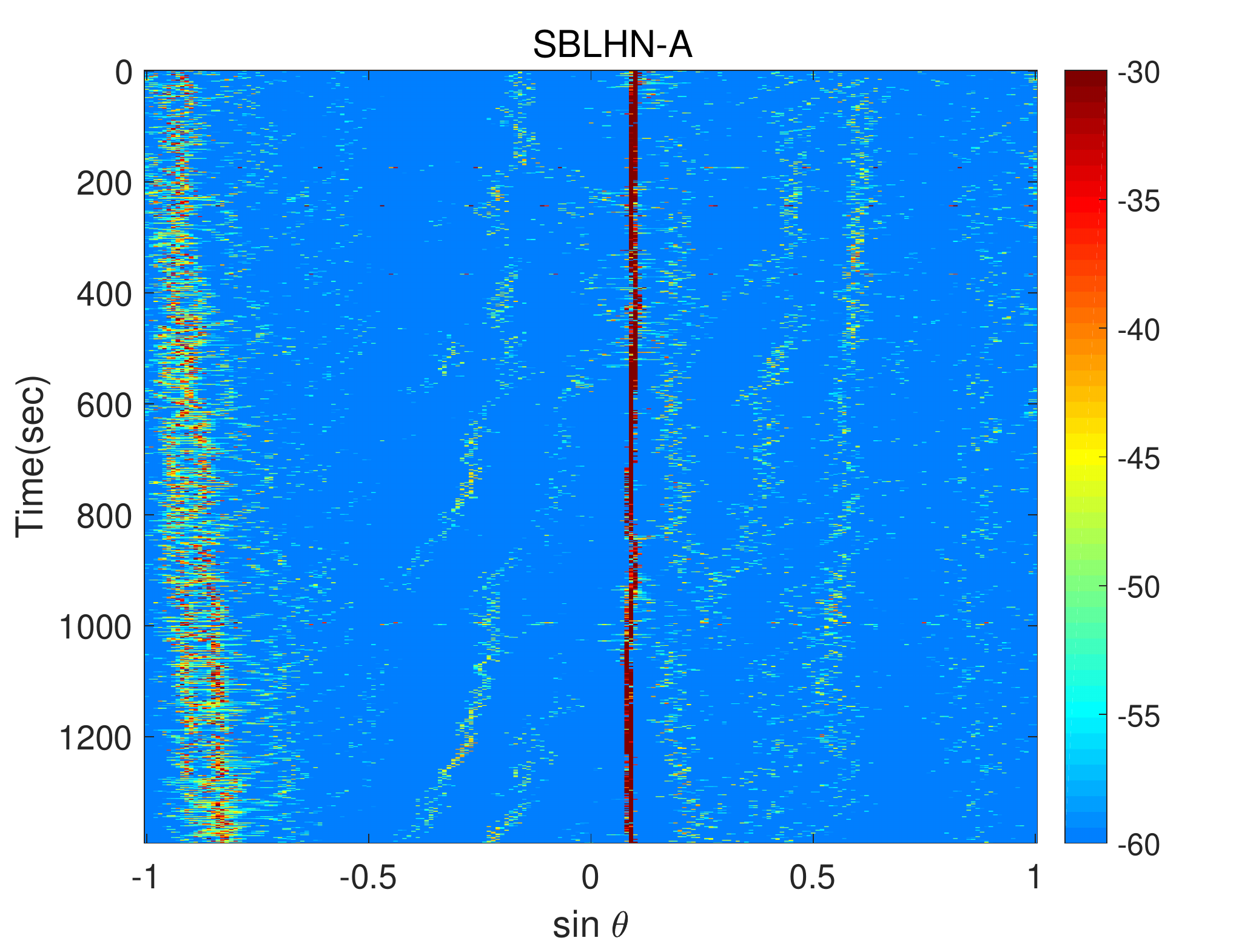}}
  \subfigure[]{
    \label{RealData3}
    \includegraphics[width=80mm]{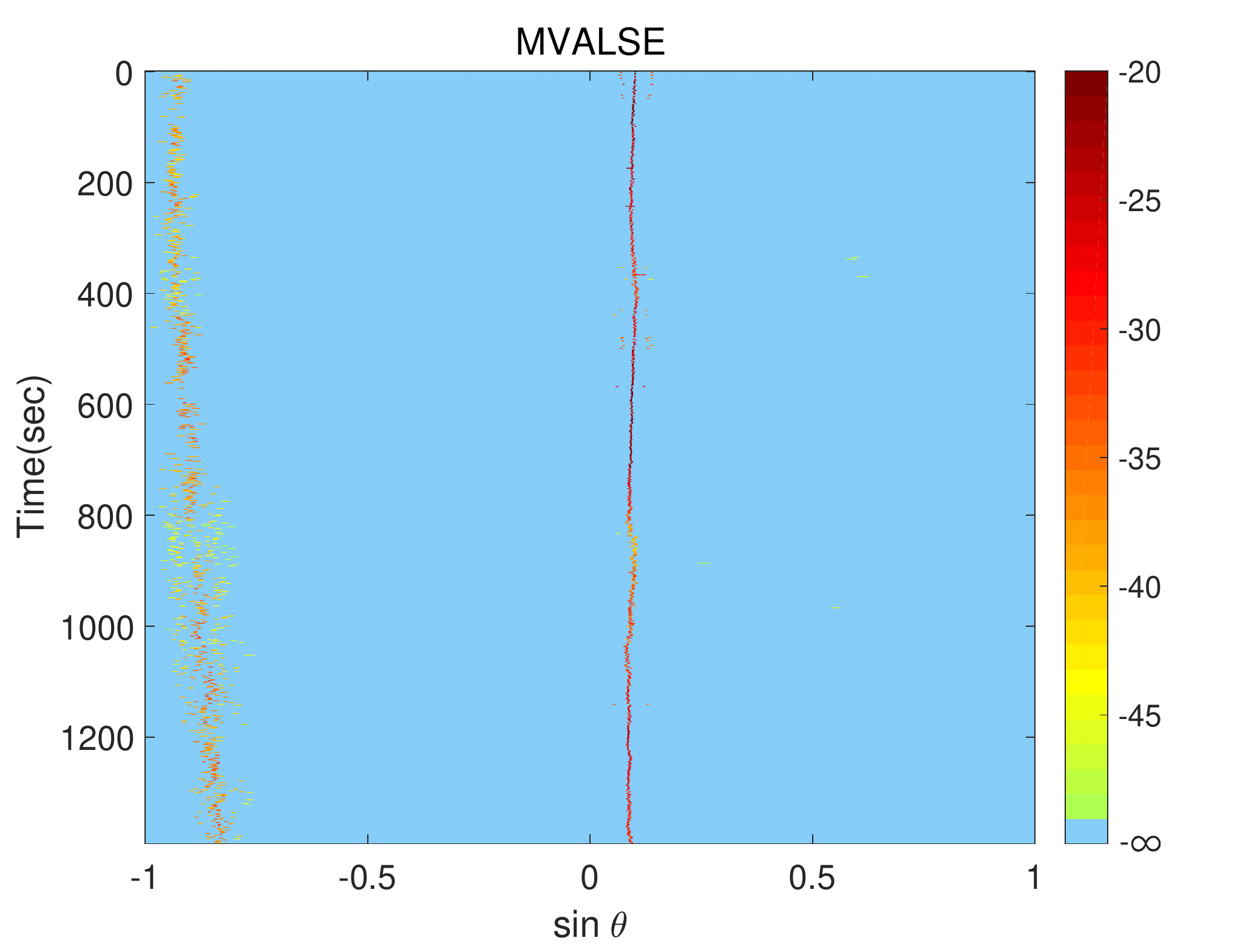}}
  \subfigure[]{
    \label{RealData4}
    \includegraphics[width=80mm]{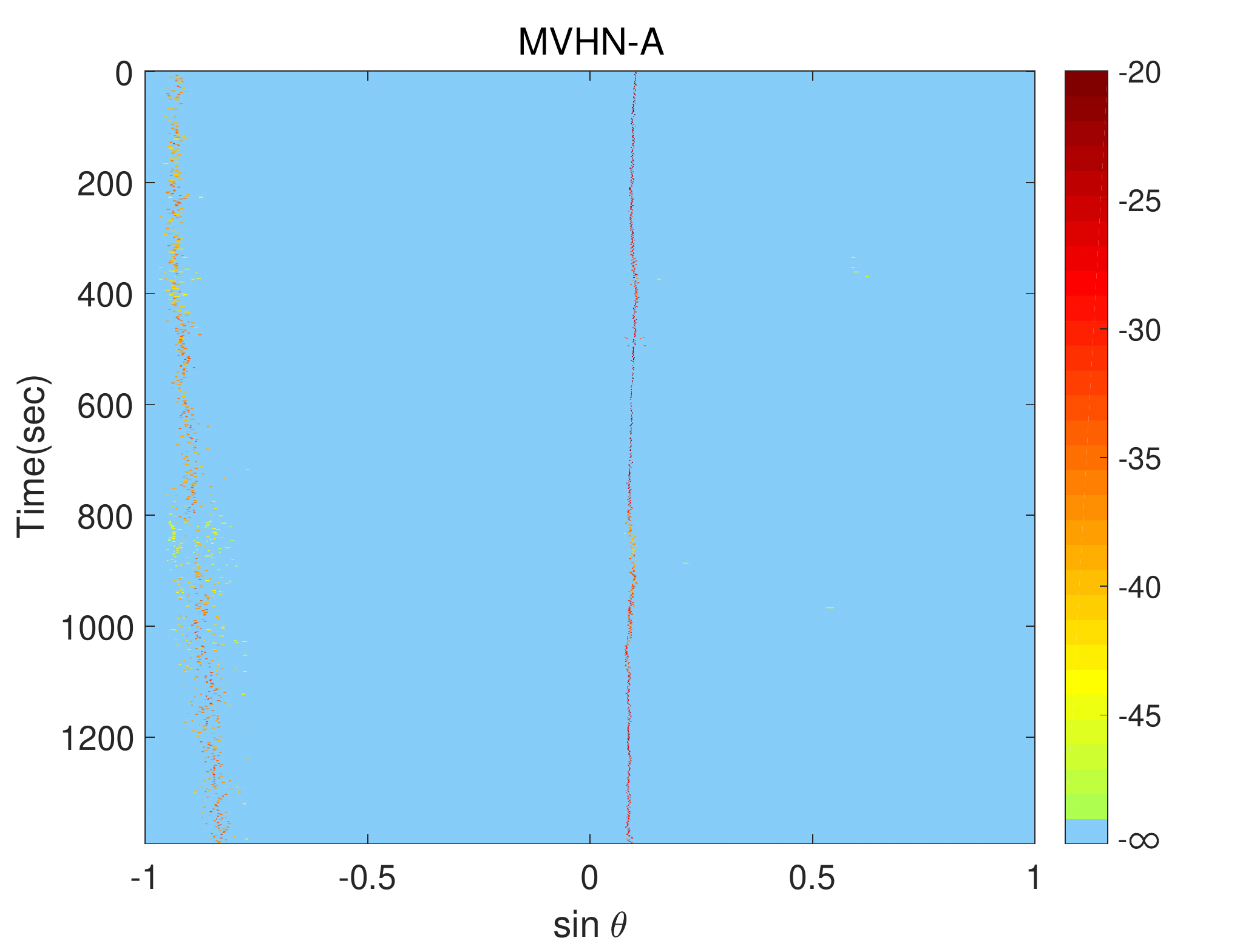}}
  \caption{The performance of algorithms for real data. (a) Beam power of CBF. (b) Signal weights of SBL. (c) Signal weights of MVALSE. (d) Signal weights of MVHN-A. }
  \label{RealData}
\end{figure*}

For CBF, the beam power is evaluated. For SBLHN-A, MVALSE and MVHN-A, the signal weights are estimated. Note that MVALSE and MVHN-A output the posterior PDFs and the estimated signal weight, and here the mean estimates plus the $\pm3$ standard deviations through the posterior PDF are plotted, as used in Sec. \ref{VSCBF}. Fig. \ref{RealData1} shows that the towed source signal is at $\sin\theta\approx 0.1$ and there is a dynamic signal near the endfire direction ($\sin \theta= -0.8\sim -0.9$) which is assumed to be a moving ship \cite{TCYang}. Besides, the towed source signal level is much stronger than the ship signal using CBF. From Fig. \ref{RealData2}-\ref{RealData4}, MVHN-A outputs the narrowest width of the bin, followed by MVALSE, SBLHN-A. This may imply that the noise is more likely to be varying across antennas. It is also worth noting that from the color bar, the maximum signal weights of MVHN-A and MVALSE are about $-20$ dB, which are higher than that of SBLHN-A (about $-30$ dB). The reason is that the SBLHN-A diffuses the energy into other spatial directions.
\section{Conclusion}
In this paper, the DOA under heteroscedastic noise is studied. The MVHN and its variants MVHN-A, MVHN-S are proposed and derived in a unified way. It is numerically shown that when the noise is wide-sense stationary only in antennas and noise is wide-sense stationary only in snapshots, MVHN-A and MVHN-S are preferred, respectively. When the noise is heteroscedastic across both antennas and snapshots, MVHN should also be avoided as its estimates are very biased. Besides, the high resolution and low complexity advantages of the proposed methods are also demonstrated, compared to the CBF and SBL based approaches. Finally, a real data set is also used to demonstrate the effectiveness of the proposed algorithm.
\section{Appendix}
\subsection{Finding a Local Maximum of $\ln Z({\mathbf s})$}\label{modelSelection}
Finding the globally optimal binary sequence $\mathbf s$ of (\ref{lnZ}) is hard in general. As a result, a greedy iterative search strategy is adopted. We proceed as follows: In the $p$th iteration, we obtain the $k$th test sequence $\mathbf t_k$ by flipping the $k$th element of $\mathbf s^{(p)}$. Then we calculate $\Delta^{(p)}_k=\ln Z(\mathbf t_k)-\ln Z({\mathbf s}^{(p)})$ for each $k=1,\cdots,N$. If $\Delta^{(p)}_k<0$ holds for all $k$ we terminate the algorithm and set $\widehat{\mathbf s}={\mathbf s}^{(p)}$, else we choose the $t_k$ corresponding to the maximum $\Delta^{(p)}_k$ as $\mathbf s^{(p+1)}$ in the next iteration.

When $k\not\in{\mathcal S}$, that is, $s_k=0$, we activate the $k$th component of $\mathbf s$ by setting $s_k^{'}=1$. Now, ${\mathcal S}'=\mathcal S\cup\{k\}$.
\begin{small}
\begin{align}\notag\label{delta_z1}
&\Delta_k = \ln Z(\mathbf s')-\ln Z(\mathbf s)\\
=&\sum_{l=1}^L\left(\ln\det([{\mathbf J}_l]_{{\mathcal S}}+\frac{1}{\tau}\mathbf I_{|\mathcal S|})-\ln\det([{\mathbf J}_l]_{{\mathcal S'}}+\frac{1}{\tau}\mathbf I_{|\mathcal S'|})\right)\notag\\
+&\sum_{l=1}^L\left({\mathbf H}_{\mathcal S',l}^{\rm H}([{\mathbf J}_l]_{{\mathcal S'}}+\frac{1}{\tau}{\mathbf I}_{|\mathcal S'|})^{-1}{\mathbf H}_{{\mathcal S'},l}-{\mathbf H}_{{\mathcal S},l}^{\rm H}([{\mathbf J}_l]_{{\mathcal S}}+\frac{1}{\tau}\mathbf I_{|\mathcal S|})^{-1}{\mathbf H}_{{\mathcal S},l}\right)\notag\\
+&L\ln\frac{1}{\tau}+\ln\frac{\rho}{1-\rho}.
\end{align}
\end{small}
By using the block-matrix determinant formula, one has
\begin{align}\label{det_J}
&\ln\det([{\mathbf J}_l]_{{\mathcal S'}}+\frac{1}{\tau}\mathbf I_{|\mathcal S'|})=\ln{\det([{\mathbf J}_l]_{{\mathcal S}}+\frac{1}{\tau}\mathbf I_{|\mathcal S|})}+{\rm ln}{\left({\rm tr}(\boldsymbol\Sigma^{-1}_l)+\frac{1}{\tau}-[{\mathbf J}_l]_{\mathcal S,k}^{\rm H}([{\mathbf J}_l]_{{\mathcal S}}+\frac{1}{\tau}{\mathbf I}_{|\mathcal S|})^{-1}[{\mathbf J}_l]_{\mathcal S,k}\right)}.
\end{align}
By the block-wise matrix inversion formula, one has
\begin{align}\label{det_H}
&\mathbf H_{{\mathcal S'},l}^{\rm H}([{\mathbf J}_l]_{{\mathcal S'}}+\frac{1}{\tau}\mathbf I_{|\mathcal S'|})^{-1}\mathbf H_{{\mathcal S'},l}
= \mathbf H_{{\mathcal S},l}^{\rm H}([{\mathbf J}_l]_{{\mathcal S}}+\frac{1}{\tau}\mathbf I_{|\mathcal S|})^{-1}{\mathbf H}_{{\mathcal S},l}+\frac{|u_{k,l}|^2}{v_{k,l}},
\end{align}
where
\begin{align}\label{v-k-u-k}
&v_{k,l} = \left({\rm tr}(\boldsymbol\Sigma^{-1}_l)+\frac{1}{\tau}-[{\mathbf J}_l]^{\rm H}_{{\mathcal S},k}([{\mathbf J}_l]_{{\mathcal S}}+\frac{1}{\tau}\mathbf I_{|\mathcal S|})^{-1}[{\mathbf J}_l]_{{\mathcal S},k}\right)^{-1},\notag\\
&u_{k,l} = v_{k,l}\left({h}_{k,l}-[{\mathbf J}_l]^{\rm H}_{{\mathcal S},k}([{\mathbf J}_l]_{{\mathcal S}}+\frac{1}{\tau}\mathbf I_{|\mathcal S|})^{-1}{\mathbf H}_{\mathcal S,l}\right).
\end{align}
Inserting (\ref{det_J}) and (\ref{det_H}) into (\ref{delta_z1}), $\Delta_k$ can be simplified as
\begin{align}\label{delta-k-active}
\Delta_k = \sum_{l=1}^L\left(\ln\frac{v_{k,l}}{\tau} + \frac{|u_{k,l}|^2}{v_{k,l}}\right)+\ln\frac{\rho}{1-\rho}.
\end{align}
Given that $\mathbf s$ is changed into ${\mathbf s}'$, the mean $\widehat{\mathbf X}'_{{\mathcal S'},l}$ and covariance $[{\widehat{\mathbf C}}'_l]_{\mathcal S'}$ of the $l$th snapshot can be updated from (\ref{W-C-1}), i.e.,
\begin{subequations}
\begin{align}\label{cov_update}
[{\widehat{\mathbf C}}'_l]_{\mathcal S'} &= ([{\mathbf J}_l]_{\mathcal S'}+\frac{1}{\tau}\mathbf I_{|\mathcal S'|})^{-1},\\
\widehat{\mathbf X}'_{{\mathcal S'},l} &= [\widehat{\mathbf C}'_l]_{{\mathcal S'}}\mathbf H_{{\mathcal S'},l}.
\end{align}
\end{subequations}
In fact, the matrix inversion can be avoided when updating $\widehat{\mathbf X}'_{{\mathcal S'},l}$ and $[{\widehat{\mathbf C}}'_l]_{\mathcal S'}$. It can be shown that
\begin{align}\label{C_active}
&\begin{bmatrix}
[{\widehat{\mathbf C}}'_l]_{\mathcal S} & [{\widehat{\mathbf C}}'_l]_{\mathcal S,k} \\
[{\widehat{\mathbf C}}'_l]_{k,\mathcal S} & [{\widehat{\mathbf C}}'_l]_{k,k}
\end{bmatrix}=
\begin{bmatrix}
[{\mathbf J}_l]_{\mathcal S}+\frac{1}{\tau}\mathbf I_{|\mathcal S|} & [{\mathbf J}_l]_{{\mathcal S},k} \\
[{\mathbf J}_l]_{k,{\mathcal S}} & {\rm tr}(\boldsymbol\Sigma^{-1}_l)+\frac{1}{\tau}\\
\end{bmatrix}^{-1}\notag\\
&=\begin{bmatrix}
 [\widehat{\mathbf C}_l]_{{\mathcal S}}^{-1} & [{\mathbf J}_l]_{{\mathcal S},k} \\
[{\mathbf J}_l]_{k,{\mathcal S}} & {\rm tr}(\boldsymbol\Sigma^{-1}_l)+\frac{1}{\tau}\\
\end{bmatrix}^{-1}=\begin{bmatrix}
[{\widehat{\mathbf C}}_l]_{\mathcal S}+v_{k,l}[{\widehat{\mathbf C}}_l]_{\mathcal S}[{\mathbf J}_l]_{\mathcal S,k}[{\mathbf J}_l]_{k,\mathcal S}[\widehat{\mathbf C}_l]_{\mathcal S}  &-v_{k,l}[\widehat{\mathbf C}_l]_{\mathcal S}[{\mathbf J}_l]_{\mathcal S,k} \\
-v_{k,l}[{\mathbf J}_l]_{k,\mathcal S}[\widehat{\mathbf C}_l]_{\mathcal S} & v_{k,l}
\end{bmatrix}.
\end{align}
Furthermore, the weight ${\widehat{\mathbf X}}'_{{\mathcal S'},l}$ is updated as
\begin{small}
\begin{align}
\begin{bmatrix}
{\widehat{\mathbf X}'_{\mathcal S,l}} \\
{\widehat{x}'_{k,l}}
\end{bmatrix}=
\begin{bmatrix}
[\widehat{\mathbf C}'_l]_{{\mathcal S}} & [{\widehat{\mathbf C}}'_l]_{{\mathcal S},k} \\
[\widehat{\mathbf C}'_l]_{k,{\mathcal S}} & [\widehat{\mathbf C}'_l]_{k,k}
\end{bmatrix}
\begin{bmatrix}
{{\mathbf H}_{\mathcal S,l}} \\
{h_{k,l}}
\end{bmatrix} =
\begin{bmatrix}
{\widehat{\mathbf X}}_{{\mathcal S},l} - [\widehat{\mathbf C}_l]_{\mathcal S}[{\mathbf J}_l]_{\mathcal S,k}u_{k,l}\\
u_{k,l}
\end{bmatrix}.\notag
\end{align}
\end{small}
For the deactive case with ${\mathbf s}_k=1$, ${\mathbf s}'_k = 0$ and ${\mathcal S}'=\mathcal S\backslash\{k\}$, $\Delta_k = \ln Z(\mathbf s')-\ln Z(\mathbf s)$ is the negative of (\ref{delta-k-active}), i.e.,
\begin{align}\label{delta-k-deactive}
\Delta_k =- \sum_{l=1}^L\left(\ln\frac{v_{k,l}}{\tau} + \frac{|u_{k,l}|^2}{v_{k,l}}\right)-\ln\frac{\rho}{1-\rho}.
\end{align}
Similar to (\ref{C_active}), the posterior mean and covariance update equation from ${\mathcal S}'$ to ${\mathcal S}$ case of $l$th snapshot can be rewritten as
\begin{small}
\begin{align}\label{C_deactive}
\begin{bmatrix}
[\widehat{\mathbf C}'_l]_{{\mathcal S}'}+v_{k,l}[\widehat{\mathbf C}'_l]_{{\mathcal S}'}[{\mathbf J}_l]_{{\mathcal S}',k}[{\mathbf J}_l]_{k,{\mathcal S}'}[\widehat{\mathbf C}'_l]_{{\mathcal S}'} & -v_{k,l}[\widehat{\mathbf C}'_l]_{{\mathcal S}'}[{\mathbf J}_l]_{{\mathcal S}',k} \\
-v_{k,l}[{\mathbf J}_l]_{k,{\mathcal S}'}[\widehat{\mathbf C}'_l]_{{\mathcal S}'} & v_{k,l}
\end{bmatrix} =\begin{bmatrix}
[\widehat{\mathbf C}_l]_{{\mathcal S}'} & [\widehat{\mathbf C}_l]_{{\mathcal S}',k} \\
[\widehat{\mathbf C}_l]_{k,{\mathcal S}'} & [\widehat{\mathbf C}_l]_{k,k}
\end{bmatrix}
\end{align}
\end{small}
\begin{align}\label{W_deactive}
\begin{bmatrix}
{\widehat{\mathbf X}'}_{{\mathcal S'},l} - [\widehat{\mathbf C}'_l]_{{\mathcal S}'}[{\mathbf J}_l]_{{\mathcal S}',k}u_{k,l}\\
u_{k,l}
\end{bmatrix}=
\begin{bmatrix}
{\widehat{\mathbf X}_{{\mathcal S}',l}} \\
{\widehat{x}_{k,l}}
\end{bmatrix},
\end{align}
According to (\ref{C_deactive}) and (\ref{W_deactive}), one has
\begin{subequations}\label{C_de_appro}
\begin{align}
[\widehat{\mathbf C}'_l]_{{\mathcal S}'}+v_{k,l}[\widehat{\mathbf C}'_l]_{{\mathcal S}'}[{\mathbf J}_l]_{{\mathcal S}',k}[{\mathbf J}_l]_{k,{\mathcal S}'}[\widehat{\mathbf C}'_l]_{{\mathcal S}'} &= [\widehat{\mathbf C}_l]_{{\mathcal S}'},\label{C_de_approa}\\
-v_{k,l}[\widehat{\mathbf C}'_l]_{{\mathcal S}'}[{\mathbf J}_l]_{{\mathcal S}',k} &= [\widehat{\mathbf C}_l]_{{\mathcal S}',k}\label{C_de_approb}\\
v_{k,l} &= [\widehat{\mathbf C}_l]_{k,k},\label{C_de_approc}\\
{\widehat{\mathbf X}'}_{{\mathcal S'},l} - [\widehat{\mathbf C}'_l]_{{\mathcal S}'}[{\mathbf J}_l]_{{\mathcal S}',k}u_{k,l}&={\widehat{\mathbf X}_{{\mathcal S}',l}},\label{C_de_approd}\\
u_{k,l}&={\widehat{x}_{k,l}}.\label{C_de_approe}
\end{align}
\end{subequations}
Thus, ${\widehat{\mathbf C}'}_{{\mathcal S}',{\mathcal S}',l}$ can be updated by substituting (\ref{C_de_approb}) and (\ref{C_de_approc}) in (\ref{C_de_approa}), i.e.,
\begin{align}
[\widehat{\mathbf C}'_l]_{{\mathcal S}'} = [\widehat{\mathbf C}_l]_{{\mathcal S}'} - \frac{[\widehat{\mathbf C}_l]_{{\mathcal S}',k}[\widehat{\mathbf C}_l]_{k,{\mathcal S}'}}{[\widehat{\mathbf C}_l]_{k,k}}.
\end{align}
Similarly, ${\widehat{\mathbf X}'}_{{\mathcal S'},l}$ can be updated by substituting (\ref{C_de_approb}) and (\ref{C_de_approe}) in (\ref{C_de_approd}), i.e.,
\begin{align}
{\widehat{\mathbf X}'}_{{\mathcal S'},l} = [\widehat{\mathbf C}'_l]_{{\mathcal S}'}[{\mathbf J}_l]_{{\mathcal S}',k}u_{k,l} + {\widehat{\mathbf X}_{{\mathcal S}',l}} = {\widehat{\mathbf X}_{{\mathcal S}',l}} -  \frac{[\widehat{\mathbf C}_l]_{{\mathcal S}',k}}{[\widehat{\mathbf C}_l]_{k,k}}\widehat{x}_{k,l}.
\end{align}
According to $v_{k,l} = [\widehat{\mathbf C}_l]_{k,k}$ (\ref{C_de_approc}) and ${u}_{k,l} = {\widehat{x}_{k,l}}$ (\ref{C_de_approe}), $\Delta_k$ (\ref{delta-k-deactive}) can be simplified as
\begin{align}\label{delta-k-deactivesim}
\Delta_k =- \sum_{l=1}^L\left(\ln\frac{{[\widehat{\mathbf C}_l]_{k,k}}}{\tau} + \frac{|{\widehat{x}_{k,l}}|^2}{{[\widehat{\mathbf C}_l]_{k,k}}}\right)-\ln\frac{\rho}{1-\rho}.
\end{align}
\subsection{The Derivation of CRB}\label{DerivationCRB}
Here we calculate the CRB for the general Case $\rm\uppercase\expandafter{\romannumeral4}$, then we specialize it to the other three Cases\footnote{The CRB under measurement incomplete scenario is straightforward and is omitted here for brevity.}. Let $g_{k,l}$ and $\phi_{k,l}$ be the amplitude and phase of $x_{k,l}$, $\forall k = 1, \cdots, K, l = 1, \cdots, L$, i.e., $g_{k,l}=|x_{k,l}|$ and $\phi_{k,l}=\angle{x_{k,l}}$. Then we obtain matrices $\mathbf G$ and $\boldsymbol \Phi$. Let $\boldsymbol \kappa$ be ${\boldsymbol \kappa}=\left[\boldsymbol \omega^{\rm T}, {\mathbf g}^{\rm T}, \boldsymbol{\phi}^{\rm T} \right]^{\rm T}$, where ${\mathbf g}={\rm vec}({\mathbf G})$ and $\boldsymbol{\phi}={\rm vec}({\boldsymbol \Phi})$. Then the Fisher information matrix (FIM) is calculated according to \cite{Fu, Lin}
\begin{align}\label{unqFIM}
{\mathbf I}({\boldsymbol \kappa})=\sum\limits_{m=1}^M \sum\limits_{l=1}^L  \frac{2}{\nu_{m,l}}\left(\frac{\partial \Re\{Z_{m,l}\}}{\partial {\boldsymbol \kappa}}\left(\frac{\partial \Re\{Z_{m,l}\}}{\partial {\boldsymbol \kappa}}\right)^{\rm T}
+\frac{\partial \Im\{Z_{m,l}\}}{\partial {\boldsymbol \kappa}}\left(\frac{\partial \Im\{Z_{m,l}\}}{\partial {\boldsymbol \kappa}}\right)^{\rm T}\right).
\end{align}
By defining ${\mathbf g}_{l}=[g_{1,l},\cdots,g_{K,l}]^{\rm T}$ and $\phi_{l}=[\phi_{1,l},\cdots,\phi_{K,l}]^{\rm T}$, we have
\begin{align}\label{vecgrad}
\frac{\partial \Re\{Z_{m,l}\}}{\partial {\boldsymbol \kappa}}=
\left[\begin{array}{c}
\frac{\partial \Re\{Z_{m,l}\}}{\partial {\boldsymbol \omega}} \\
{\mathbf 0}_{(l-1)K} \\
\frac{\partial \Re\{Z_{m,l}\}}{\partial {\mathbf g}_{l}} \\
{\mathbf 0}_{(L-l)K} \\
{\mathbf 0}_{(l-1)K} \\
\frac{\partial \Re\{Z_{m,l}\}}{\partial {\boldsymbol \phi}_{l}} \\
{\mathbf 0}_{(L-l)K} \\
\end{array}\right]
,~\frac{\partial \Im\{Z_{m,l}\}}{\partial {\boldsymbol \kappa}}=
\left[\begin{array}{c}
\frac{\partial \Im\{Z_{m,l}\}}{\partial {\boldsymbol \omega}} \\
{\mathbf 0}_{(l-1)K} \\
\frac{\partial \Im\{Z_{m,l}\}}{\partial {\mathbf g}_{l}} \\
{\mathbf 0}_{(L-l)K} \\
{\mathbf 0}_{(l-1)K} \\
\frac{\partial \Im\{Z_{m,l}\}}{\partial {\boldsymbol \phi}_{l}} \\
{\mathbf 0}_{(L-l)K} \\
\end{array}\right],
\end{align}
where
\begin{subequations}
\begin{align}
&\frac{\partial \Re\{Z_{m,l}\}}{\partial {\omega_k}} = -(m-1)g_{k,l}{\rm {sin}}\left[(m-1)\omega_k + \phi_{k,l} \right],\notag\\
&\frac{\partial \Re\{Z_{m,l}\}}{\partial {g_{k,l}}} = {\rm {cos}}\left[(m-1)\omega_k + \phi_{k,l} \right],\notag\\
&\frac{\partial \Re\{Z_{m,l}\}}{\partial {\phi_{k,l}}} = -g_{k,l}{\rm {sin}}\left[(m-1)\omega_k + \phi_{k,l} \right],\notag\\
&\frac{\partial \Im\{Z_{m,l}\}}{\partial {\omega_k}} = (m-1)g_{k,l}{\rm {cos}}\left[(m-1)\omega_k + \phi_{k,l} \right],\notag\\
&\frac{\partial \Im\{Z_{m,l}\}}{\partial {g_{k,l}}} = {\rm {sin}}\left[(m-1)\omega_k + \phi_{k,l} \right],\notag\\
&\frac{\partial \Im\{Z_{m,l}\}}{\partial {\phi_{k,l}}} = g_{k,l}{\rm {cos}}\left[(m-1)\omega_k + \phi_{k,l} \right].\notag
\end{align}
\end{subequations}
Substituting (\ref{vecgrad}) in (\ref{unqFIM}), the FIM ${\mathbf I}({\boldsymbol \kappa})$ is obtained. The CRB is ${\rm CRB}({\boldsymbol \kappa})={\mathbf I}^{-1}({\boldsymbol \kappa})$ and CRB of frequencies are $[{\rm CRB}({\boldsymbol \kappa})]_{1:K,1:K}$, which will be used as the performance metrics.

Substituting $\nu_{m,l}=\nu$, $\forall m,l$, $\nu_{m,l}=\nu_m$, $\forall l$, $\nu_{m,l}=\nu_l$, $\forall m$ in (\ref{unqFIM}), we obtain the FIM for Case $\rm \uppercase\expandafter{\romannumeral1}$, $\rm \uppercase\expandafter{\romannumeral2}$, $\rm \uppercase\expandafter{\romannumeral3}$, respectively. Taking the inverse of the FIM yields the CRB.
\section{Acknowledgement}
The authors would like to thank T. C. Yang for valuable discussions and suggestions on this work.
\bibliographystyle{IEEEbib}
\bibliography{strings,refs}

\end{document}